\documentclass{aa}  
\usepackage{natbib}
\usepackage[colorlinks=true,citecolor=blue]{hyperref}
\usepackage{soul}
\usepackage{xcolor}
\usepackage[page]{appendix}
\usepackage{upgreek}
\bibpunct{(}{)}{;}{a}{}{,}
\usepackage{graphicx}
\usepackage{txfonts,textcomp}
\newcommand{\bpic}[1]{$\upbeta$\,Pictoris}
\newcommand{\HD}[1]{HD\,110058}

\newcommand{\micron}[1]{\,$\upmu$m}

\begin{document} 

\title{An inner warp discovered in the disk around \HD{} using VLT/SPHERE\thanks{Based on observations made with ESO telescopes at the Paranal Observatory under programmes 095.C-0389(A) and ID 095.C-0607(A)} and HST/STIS\thanks{Based on observations made with the STIS instrument aboard HST under program GO-15218 supported by NASA}}

\author{S.~Stasevic\inst{1,2}, J.~Milli\inst{1}, J.~Mazoyer \inst{2}, A.-M.~Lagrange\inst{1,2}, M.~Bonnefoy\inst{1}, V.~Faramaz-Gorka\inst{3}, F.~M\'{e}nard\inst{1}, A.~Boccaletti\inst{2}, E.~Choquet\inst{4}, L.~Shuai\inst{7}, J.~Olofsson\inst{6}, A.~Chomez\inst{1,2}, B.~Ren\inst{5}, P.~Rubini\inst{8}, C.~Desgrange\inst{1,6}, R.~Gratton\inst{9}, G.~Chauvin\inst{1,5}, A.~Vigan\inst{4}, E.~Matthews\inst{6}}

\institute{
Univ. Grenoble Alpes, CNRS, IPAG, F-38000 Grenoble, France  
\and 
LESIA, Observatoire de Paris, Universit\'{e} PSL, Sorbonne, Universit\'{e}, Universit\'{e} Paris Cit\'{e}, CNRS, 5 place Jules Janssen, 92195 Meudon, France
\and 
Steward Observatory, Department of Astronomy, University of Arizona, 933 N.~Cherry Ave, Tucson, AZ 85721, USA
\and
Aix-Marseille Univ., CNRS, CNES, LAM, 38 rue Frédéric Joliot-Curie, 13388 Marseille Cedex 13, France
\and
Université Côte d'Azur, Observatoire de la Côte d'Azur, CNRS, Laboratoire Lagrange, France
\and
Max Planck Institut f\"{u}r Astronomie, K\"{o}nigstuhl 17, D-69117 Heidelberg, Germany
\and 
Department of Astronomy, Xiamen University, 1 Zengcuoan West Road, Xiamen, Fujian 361005, People's Republic of China
\and
pascal.rubini@pixyl.io 5 Avenue du Grand Sablon, 38700 La Tronche, France
\and
INAF-Osservatorio Astronomico di Padova, Vicolo dell'Osservatorio 5, Padova, Italy, 35122-I
}

\date{}%Received ; accepted 

%
%-------------------------------------------------------------------
% \abstract{}{}{}{}{} 
% 5 {} token are mandatory
 
\abstract
   {An edge-on debris disk was detected in 2015 around the young, nearby A0V star \HD{}. The disk showed features resembling those seen in the disk of \bpic{} that could indicate the presence of a perturbing planetary-mass companion in the system.}
   {We investigated new and archival scattered light images of the disk in order to characterise its morphology and spectrum. In particular, we analysed the disk's warp to constrain the properties of possible planetary perturbers.}
   {Using data from two VLT/SPHERE observations taken with the Integral Field Spectrograph (IFS) and near InfraRed Dual-band Imager and Spectrograph (IRDIS), we obtained high-contrast images of the edge-on disk. Additionally, we used archival data from HST/STIS with a poorer inner-working angle but a higher sensitivity to detect the outer parts of the disk. We measured the morphology of the disk by analysing vertical profiles along the length of the disk to extract the centroid spine position and vertical height. We extracted the surface brightness and reflectance spectrum of the disk.}
   {We detect the disk between 20\,au (with SPHERE) and 150\,au (with STIS), at a position angle of $159.6^\circ \pm 0.6^\circ$. Analysis of the spine shows an asymmetry between the two sides of the disk, with a $3.4^\circ \pm 0.9^\circ$ warp between $\sim$\,20\,au and 60\,au. The disk is marginally vertically resolved in scattered light, with a vertical aspect ratio of $9.3 \pm 0.7\%$ at 45\,au. The extracted reflectance spectrum is featureless, flat between 0.95\micron{} and 1.1\micron{}, and red from 1.1\micron{} to 1.65\micron{}. The outer parts of the disk are also asymmetric with a tilt between the two sides compatible with a disk made of forward-scattering particles and seen not perfectly edge-on, suggesting an inclination of $<84^\circ$.}
   {The presence of an undetected planetary-mass companion on an inclined orbit with respect to the disk could explain the warp. The misalignment of the inner parts of the disk with respect to the outer disk suggests a warp that has not yet propagated to the outer parts of the disk, favouring the scenario of an inner perturber as the origin of the warp.}


\authorrunning{S.~Stasevic et al.}
%\titlerunning{An inner warp discovered in the disk around \HD{}} %only for referee format
\maketitle
%
%-------------------------------------------------------------------
\section{Introduction}

%- Brief definition of debris disks, direct imaging of disks; interest of studying their morphology, planet-disk interactions; examples of asymmetric disks.
Initially detected as an infrared excess of the host star, advancements in telescope instruments for high-contrast imaging have made the direct detection of debris disks possible, with over fifty now having been resolved in scattered light \citep{Esposito2020, Xie2022}. Unlike the primordial, planet-forming disks that form during the pre-main sequence stage of a star, debris disks are composed of short-lived dust particles down to a micrometre in size, detected around main sequence stars, and sustained through continuous collisions of larger bodies in a steady-state equilibrium \citep{Hughes2018}. Many mechanisms act to dissipate small dust grains from the system, such as the dispersal of gas coupled to the dust via photoevaporation, stellar winds, radiation pressure, and accretion \citep{Adams2004}. Therefore, detection of debris disks older than the expected dispersal time of $\sim$\,10\,Myr \citep{Andrews2020} indicates that the system was successful in forming numerous enough planetesimals to sustain dust production over a long time period, and perhaps planetary-mass companions that stir the planetesimals. Planetary companions are capable of shaping unique structures -- such as gaps, warps, and asymmetries -- in the disk \citep{Lee2016}, and detections of such structures in directly imaged debris disks may indicate the presence of as of yet undetected planetary companions in the system. 

%- Information on HD110058 (in Sco-Cen, spectral type, distance, age, etc.)
Direct imaging surveys have achieved the sensitivity to detect planetary-mass companions for a select number of debris disk stars, including \bpic{} \citep{Lagrange2010}, HR\,8799 \citep{Marois2008, Marois2010}, HD\,95086 \citep{Rameau2013}, and HD\,106906 \citep{Bailey2014}. In the case of \bpic{} in particular, the existence of a planetary-mass companion was predicted prior to the detection of \bpic{} b from the warped inner disk seen in scattered light images \citep{Mouillet1997}. The orbit of the planet was later confirmed to be inclined from the plane of the disk \citep{Lagrange2012}, closely matching the predictions made using dynamical modelling \citep{Augereau2001}.

We performed a morphological analysis of the edge-on disk around the A0 type star \HD{} (HIP\,61782), first published by \cite{Kasper2015}, which shows structures that closely resemble the warp seen in \bpic{}. It is located at a distance of $130.08 \pm 0.53$\,pc \citep{GaiaDR3}, in the Lower Centaurus-Crux sub-group of the Scorpius-Centaurus (Sco-Cen) OB association \citep{deZeeuw1999}, which has an estimated age of 17\,Myr \citep{Pecaut2012}. The presence of CO gas in the disk has been detected using the Atacama Large Millimetre/submillimetre Array interferometer (ALMA) \citep{LS2016}, with \cite{Hales2022} showing that the gas distribution is consistent with that of millimetre dust that is secondary in origin. The fractional luminosity of the disk was found to be $1.9\times10^{-3}$ \citep{LS2016, Kral2017}. While the disk has now been resolved by multiple instruments -- for example, the Very Large Telescope (VLT)/Spectro-Polarimetric High-contrast Exoplanet REsearch (SPHERE), \cite{Kasper2015}; Gemini Planet Imager (GPI), \cite{Esposito2020}; ALMA, \cite{Hales2022}; Hubble Space Telescope (HST)/Space Telescope Imaging Spectrograph (STIS), \cite{Ren2023} -- a bound companion has yet to be detected.

%- Summary of Kasper+ 2015 discovery paper.
\cite{Kasper2015} traced the disk in scattered light between $\sim$\,200\,mas and 600\,mas ($26\mbox{--}78$\,au at a distance of 130\,pc) in the K and YJH band, and found a position angle (PA) of $155^{\circ} \pm 1^\circ$, with two symmetric areas of enhanced surface brightness at $\sim$\,300\,mas (39\,au), which is expected for an optically thin ring-like disk seen edge on. The disk was not spatially resolved in the vertical direction, suggesting a vertical scale height of $\lesssim$\,5\,au, and a near-perfect edge-on viewing angle, similar to \bpic{}. Evidence of anticlockwise warps on both sides of the disk were also found, starting from the radius of the ring at 300\,mas (39\,au), curving away by $\sim$\,$15^\circ$ on the south-eastern (SE) side, and roughly perpendicular to the disk plane on the north-western (NW) side, most prominent in the Y band. Detection sensitivity limits, combined with the non-detection of a companion, suggest that a super-Jupiter that is comparable to HR\,8799\,b, c, d, e, or \bpic{}\,b is not present at large separations; however, a \bpic{}\,b analogue at a smaller separation would remain undetectable in the data. 

%- Summary of additional analysis done for this paper + describe structure.
In this paper we present new high-contrast imaging observations obtained with the VLT/SPHERE \citep{Beuzit2019} instrument as part of a dedicated programme to image disks and planets around A--F stars of Sco-Cen \citep{2017A&A...597L...7B, 2021A&A...655A..62B} alongside the data presented in \cite{Kasper2015}. The data and methods of reduction and analysis are described in Sect.~\ref{sec:data}. We present the morphological analysis of the disk at large distances using archival optical data from HST/STIS in Sect.~\ref{sec:STIS_data}, and analysis of the near-infrared (NIR) data morphology, surface brightness, and reflectance spectrum in Sect.~\ref{sec:properties}. Section~\ref{sec:interp} covers the interpretation of our findings, including comparison to the disk of \bpic{}, as well as the recently published ALMA observations from \cite{Hales2022}, and dynamical models. Finally, we present constraints on planetary mass companions which may be responsible for the features seen in the disk in Sect.~\ref{sec:planets} before concluding in Sect.~\ref{sec:concl}.

%
%-------------------------------------------------------------------
\section{Near-infrared data}
\label{sec:data}
\subsection{Observations}
\label{subsec:obs}

We used data from two SPHERE observations of \HD{} observed on 2015 April 4 \cite[observation night 2015 April 3; data set presented in][]{Kasper2015} and 2015 April 13 (observation night 2015 April 12; new data set part of the ESO programme ID: 095.C-0607(A); PI: M.~Bonnefoy) using the infrared dual-band imager and spectrograph (IRDIS) \citep{Dohlen2008, Vigan2010} and integral field spectrograph (IFS) \citep{Claudi2008} in parallel.

Table~\ref{tab:obs} summarises the observing setup and conditions. The star was observed with the IFS from the Y to H band ($0.95\mbox{--}1.65$\micron{}; R$\sim$\,30) and Y to J band ($0.95\mbox{--}1.35$\micron{}; R$\sim$\,50) for both epochs respectively, covering a 1.7$\arcsec$\,$\times$\,1.7\arcsec{} field of view. The IRDIS observations were carried out using the K12 (centred at 2.103\micron{} and 2.254\micron{} with filter bandwidths of 0.101\micron{} and 0.110\micron{} respectively) and H23 (centred at 1.586\micron{} and 1.666\micron{} with filter bandwidths of 0.053\micron{} and 0.056\micron{} respectively) filters for both epochs respectively, covering a 11$\arcsec$\,$\times$\,11\arcsec{} field of view. 
During both observations, the star was placed behind the N\_ALC\_YJH\_S apodised Lyot coronagraph and 185\,mas wide focal mask.

Both observation sequences were conducted in pupil-stabilised mode in order to perform angular differential imaging (ADI) \citep{Marois2006}, a technique for reducing quasi-static stellar noise arising from wave front perturbations during the observation \citep{Oppenheimer2009}. The ADI reduction method utilises the field of view rotation between frames to reconstruct the diffraction and speckle pattern -- whose orientation does not change with the field of view -- while avoiding astrophysical signal. In addition to the ADI sequence and sky and flat-field exposures for performing standard calibrations, non-saturated images of the star placed outside of the coronagraphic mask were recorded to obtain a reference point-spread function (PSF) and relative photometric calibration. Coronagraphic images of the star with four satellite footprints of the PSF created by the deformable mirror were taken in order to retrieve the position of the star behind the mask.

\begin{table*}[t]
\caption{VLT/SPHERE observations of \HD{} used in this study.}
\label{tab:obs}
\begin{center}
\begin{tabular}{lllllllllll}
\hline\hline
     & \multicolumn{3}{c}{IRDIS} & \multicolumn{3}{c}{IFS} &    &   \\
        \cline{2-5}         \cline{6-9}
Observation night & Filter & DIT & Frames & $\Delta$PA & Filter & DIT & Frames & $\Delta$PA & Seeing \\
YYYY-MM-DD &     & (s) &    & ($^\circ$) &      & (s) &   & ($^\circ$) & ($\arcsec$) \\
\hline
2015-04-03 & K12 & 8 & 176 & 14.62 & YJH & 16 & 100 & 16.1 & 0.91 \\
2015-04-12 & H23 & 16 & 224 & 34.25 & YJ & 16 & 210 & 34.2 & 1.40 \\  
\hline

\end{tabular}
\end{center}
\tablefoot{The seeing is measured by the differential image motion monitor (DIMM) at 0.5\micron{}.The DIT {(detector integration time)} refers to the single exposure time, and $\Delta$PA to the amplitude of the parallactic rotation.}\\
\end{table*}

\subsection{Data reduction}
\label{subsec:reduction}

The raw data were processed with the data handling software (DRH) \citep{Pavlov2008} of the SPHERE data centre \citep{Delorme2017}. The DRH performs dark, flat, and bad pixel correction on the raw frames of the non-coronagraphic PSF and coronagraphic sequence and registers the frames. Additional wavelength calibration, bad pixel, and instrument cross-talk corrections are applied to the IFS data \citep{Mesa2015}. Astrometric calibration of the IFS and IRDIS data is performed on-sky as detailed in \cite{Maire2016}, leading to an adapted plate scale of $7.46 \pm 0.02$\,mas/pixel for the IFS, and updated values from \cite{Maire2021} for the IRDIS plate scales (H2, $12.250 \pm 0.004$; H3, $12.244 \pm 0.003$; K1, $12.258 \pm 0.004$; K2, $12.253 \pm 0.003$\,mas/pixel), and true north correction of $-1.76^\circ \pm 0.04^\circ$.

\subsubsection{Angular differential imaging}
The ADI reduction was applied on the processed data cubes following the principal component analysis (PCA) reduction method described in \cite{Soummer2012} and \cite{Amara2012} and implemented in the \emph{SpeCal} pipeline \citep{Galicher2018}. The data cubes were reshaped into 2D arrays for each wavelength channel, with one temporal and one spatial axis, and subtracted by their temporal mean prior to their eigen-decomposition \citep[adapted from][]{Gonzalez2017_VIP}. All subsequent steps of the PCA reduction were performed on non-mean centred data. The number of principal components (PCs) used was varied between 1 and 10. Reductions with a higher number of PCs proved too aggressive for preserving the extended disk structure. After subtracting the PCs from the ADI sequence, the reduced frames were derotated by their parallactic angle and median combined. Reduced frames were also derotated by the negative of their parallactic angle before being stacked in order to provide a 'diskless' image for noise estimation. 

\begin{figure*}
    \centering
    \includegraphics[width=12.9cm]{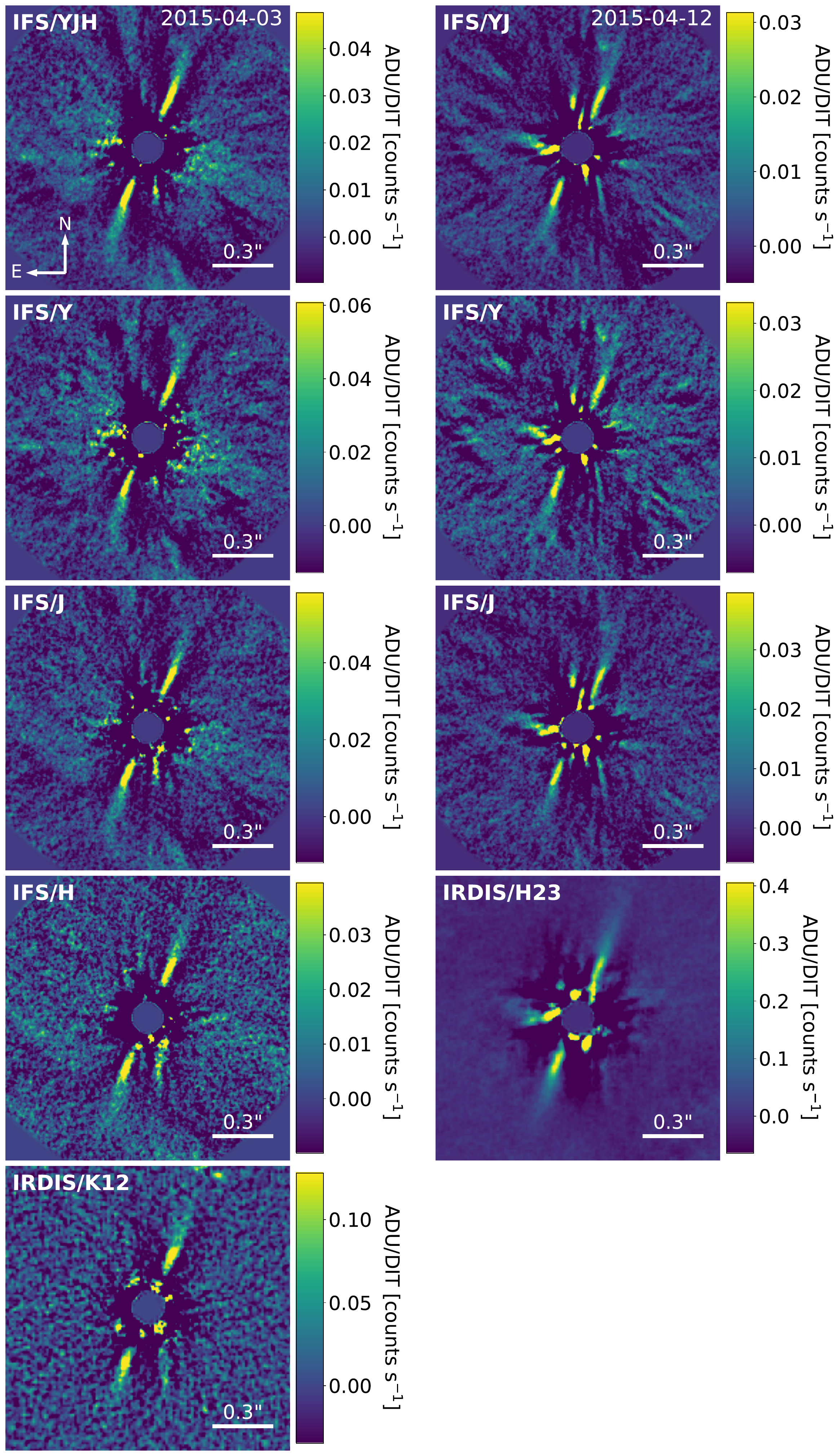}
    \caption{ADI-PCA reductions of 2015-04-03 (left) and 2015-04-12 (right) observations. The top row shows the combined image for all IFS wavelength channels, followed by rows for the Y, J; H, and K band wavelengths. The K12 band (bottom left) and 2015-04-12 H23 band (bottom right) are IRDIS, and the others IFS. Colour bars in ADU.s$^-1$ are shown to the right of each image.}
    \label{fig:ADI_img}
\end{figure*}

Signal to noise (S/N) maps were computed for each channel by subtracting the mean background flux of the reduced image, and dividing by the standard deviation of the diskless image in concentric annulii with radial widths equal to the full-width half-maximum (FWHM) of the stellar PSF in that channel. The S/N of the disk was used to weight each channel before mean combining the reduced data to produce broad-band images. For the IFS data, channels in different wavelength ranges were combined to mimic the Y, J, and H bands, using ranges of $0.95\mbox{--}1.11$\micron{}, $1.14\mbox{--}1.35$\micron{}, and $1.40\mbox{--}1.65$\micron{} respectively. Additional Y and J band stacked images were created using data from both epochs to further increase the disk S/N. The ADI-PCA reductions of the data using 5\,PCs are shown in Fig.~\ref{fig:ADI_img}. 

\subsubsection{Reference-star differential imaging}
\label{subsec:RDI}
=
In addition to the ADI reductions, the 2015 April 12 H23 IRDIS data were reduced using reference-star differential imaging (RDI, \cite{Lafreniere2009}), which uses a library of reference frames obtained from observations of other stars to compute the PCs used to reduce the science frames. While ADI reductions (especially of extended structures such as disks) suffer from self-subtraction effects where there is overlap of the physical signal between frames \citep{Milli2012}, RDI does not. Although the nearly edge-on orientation of the disk allows for minimal self-subtraction at sufficient separations with ADI, as the separation decreases, self-subtraction effects will cause the disk scale height to be underestimated. We therefore used RDI to reduce the observations for scale height measurements. 

The reference library was constructed from archival IRDIS data observed between 2014 May 13 and 2021 June 2 in the DB\_H23 filter, with the same instrument set-up as in Sect.~\ref{subsec:obs} and raw data processing as in Sect.~\ref{subsec:reduction} \cite[for a detailed review of RDI performance using SPHERE/IRDIS data, see][]{Xie2022}. The Pearson correlation coefficient was used to identify the reference frames which best matched the science data. This was computed between the target and reference frames within a circular annulus from 0.18$\arcsec{}$ to 0.49\arcsec, where the speckle noise is dominant. The frames with the highest correlation were selected for the reference library. Selection was made separately for each wavelength channel, and resulted in a final reference library consisting of 477 and 474 frames for the H2 and H3 channels respectively. 

\begin{figure*}
    \centering
    \includegraphics[width=17.5cm]{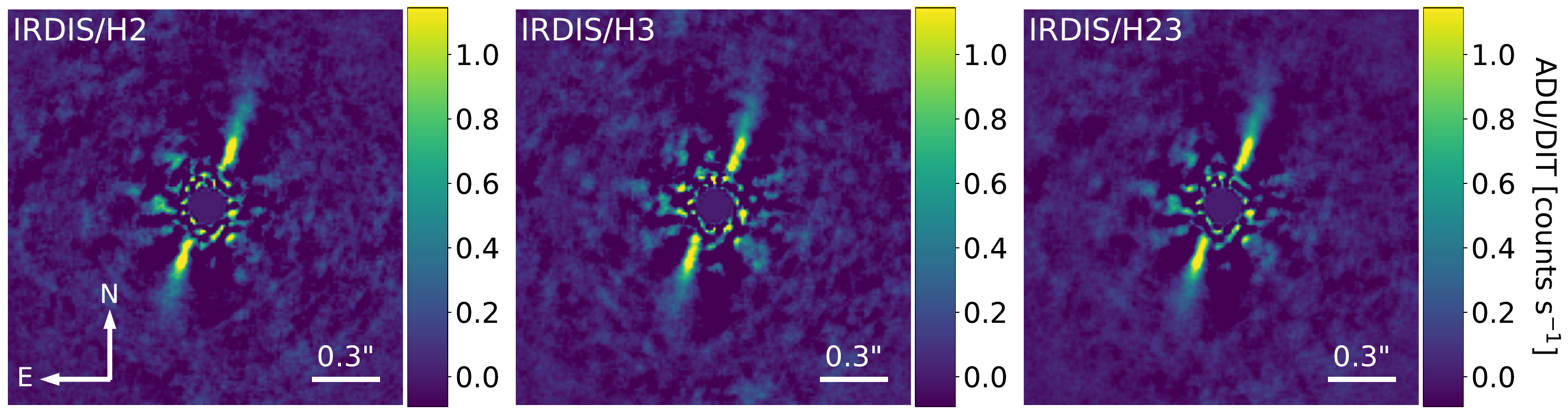}
    \caption{RDI-PCA reduction of 2015-04-12 IRDIS data using 125\,PCs for the H2 (left), H3 (centre), and combined H23 (right) channels.}
    \label{fig:RDI_img}
\end{figure*}

The PCA reduction was applied to the target observation using the reference library to construct the PCs, but otherwise followed the reduction procedure described in Sect.~\ref{subsec:reduction}. As the disk signal is not present in the reference library, a larger number of PCs can be used in the reduction, better subtracting the speckle noise. The number of PCs was varied between 50 and 200 in steps of 25. The RDI-PCA reduction using 125\,PCs (where the increase in disk S/N with PCs plateaus) is shown in Fig.~\ref{fig:RDI_img}.

\subsection{Flux loss calibration}
\label{subsec:loss}

The flux loss due to the ADI reduction was calculated in order to extract the reflectance spectrum of the disk without the assumption of achromatic self-subtraction (see Sect.~\ref{subsec:reflectance} for results). A synthetic disk with a radius of 40\,au, vertical height of 1.5\,au at that radius, and maximum flux of 1 count/s was created using the \texttt{scattered\_light\_disk} and \texttt{fakedisk} modules of the Vortex Image Processing (VIP)\footnote{\url{https://github.com/vortex-exoplanet/VIP}.} python package \citep{Gonzalez2017_VIP, Augereau1999}, convolved with the observed PSF at each wavelength, and injected into the data cube perpendicular to the real disk. The parameters used for the fake disk were estimated from the real disk (see Sect.~\ref{subsec:spine}), with assumed forward scattering and a fiducial Henyey-Greenstein phase function \citep{Hong1985}. Since the fake disk injection is only used to assess the effects of the PCA reduction, a more thorough model fit of the disk parameters was not performed, and parameter values were only selected for the fake disk to approximate the real disk. 

The ADI-PCA reduction was applied to the injected fake-disk cube as in Sect.~\ref{subsec:reduction}. The flux of both the reduced and unconvolved fake disk was measured at each wavelength using a rectangular aperture at 0.3\arcsec{} with dimensions of 0.060\arcsec\,$\times$\,0.045\arcsec, rounded to the nearest pixel after converting by the instrument plate scale. This region was chosen as it is the brightest area of -- and hence where the flux will be extracted from -- the real disk. The ratio between the two fluxes was calculated for the two sides of the disk separately in each channel. While the RDI reductions do not suffer from self-subtraction, over-subtraction can still occur, so fake disk injections of the IRDIS H23 data were also reduced using the same RDI procedure described in Sect.~\ref{subsec:RDI}.

%
%-------------------------------------------------------------------
\section{Archival optical data from HST/STIS}
\label{sec:STIS_data}

We carried out additional analysis on archival observations of the system from HST/STIS. This helps us get a broader picture of the system, as we are able to image the outer parts of the disk, complementing the view of the innermost disk that we see with VLT/SPHERE. It also allows us to more reliably measure the position angle of the unwarped component of the disk, which we used in the analysis of the NIR data.

\HD{} was observed with STIS under programme ID: HST GO-15218 (PI: \'{E}. Choquet) using the 50CORON aperture (unfiltered, with a pivot wavelength of 574\,nm and a bandwidth of 531\,nm FWHM), with a plate scale of 50.72\,mas/pixel. The target was observed on 2019 April 27 in a non-interrupted sequence including one visit to a colour-matched reference star (HD\,107800) and three visits to the science target, each with a different orientation of the telescope to maximise the azimuthal coverage of the disk around the occulter and diffraction spikes (telescope orient of $-8.9^\circ \pm 18^\circ$). Each visit included short exposures using the BAR5 occulter position (108 exposures for \HD{} for a total of 1404\,s), which provides the smallest inner working angle of the STIS instrument \citep[0.3\arcsec-wide,][]{Schneider2017}, and longer exposures using the wider WEDGEB1.0 occulter position (1\arcsec-wide, 24 exposures, total 3304.8\,s), to maximise the S/N on the fainter outer parts of the disk.

The data were processed and calibrated by the STIS calibration pipeline (\emph{calstis}), with an additional custom correction of the bad pixels (those flagged as such in the Data Quality map, as well as those identified with sigma-clipping). The starlight subtraction was performed using the classical RDI method, similarly to \citet{Schneider2014}. First, the star centre was located on the first reference frame using a Radon transform \citep{Pueyo2015}, and all the other frames (science and reference targets) were registered to that frame with a least-squares minimisation method, masking out everything in the image but the spiders beyond a radius of 20\,pixels ($\sim$\,1\arcsec). The starlight was then subtracted from each of the science images by finding the reference image and the scaling coefficient that optimises the subtraction of the spiders in that same mask. This process was done independently for all BAR5 frames and all WEDGEB1.0 frames. All frames were then rotated so  north is up, mean-combined accounting for the actual exposure time of each pixel (with or without each occulter), and calibrated to physical units (Jy.arcsec$^{-2}$) using the STIS absolute photometric calibration values (PHOTFNU keyword). The final processed image is shown in Fig.~\ref{fig:STIS_spine}. These data are also presented in \cite{Ren2023}. 

%
%-------------------------------------------------------------------
\section{Disk properties}
\label{sec:properties}

In this section we present the findings from our analysis of the different data sets. The morphology of the disk at large separations using the HST/STIS data is presented in Sect.~\ref{subsec:STIS_spine}, followed by the morphology of the disk at smaller separations, seen with the VLT/SPHERE observations, in Sect.~\ref{subsec:spine}. We then present the vertical scale height and surface brightness extracted from spine fitting in Sects.~\ref{subsec:scale_height} and \ref{subsec:s_brightness} respectively, and finally the reflectance spectrum of the disk measured from the IFS wavelength channels in Sect.~\ref{subsec:reflectance}. 

\subsection{Morphology at large separations}
\label{subsec:STIS_spine}

To determine the shape of the disk and presence of the warp, we measured the position of the disk spine, defined here as the curve starting from the star and joining the brightest pixels of the disk. The spine position was found from the centroid position of Gaussians fit to vertical profiles along the disk. Further details for the fitting process can be found in Appendix~\ref{appendix:Gauss_fit}. A least squares straight line fit was performed on the error weighted spine position of each side of the disk between 132\,au and 152\,au, where the spine appears both straight and symmetric across both sides. The slopes were converted to angles, and mean combined to give the final PA measurement of the disk. The error in each fit was calculated from its covariance matrix, and combined in quadrature between the two sides to give the final error of the mean PA measurement.

\begin{figure}[h]
    \centering
    \includegraphics[width=9cm]{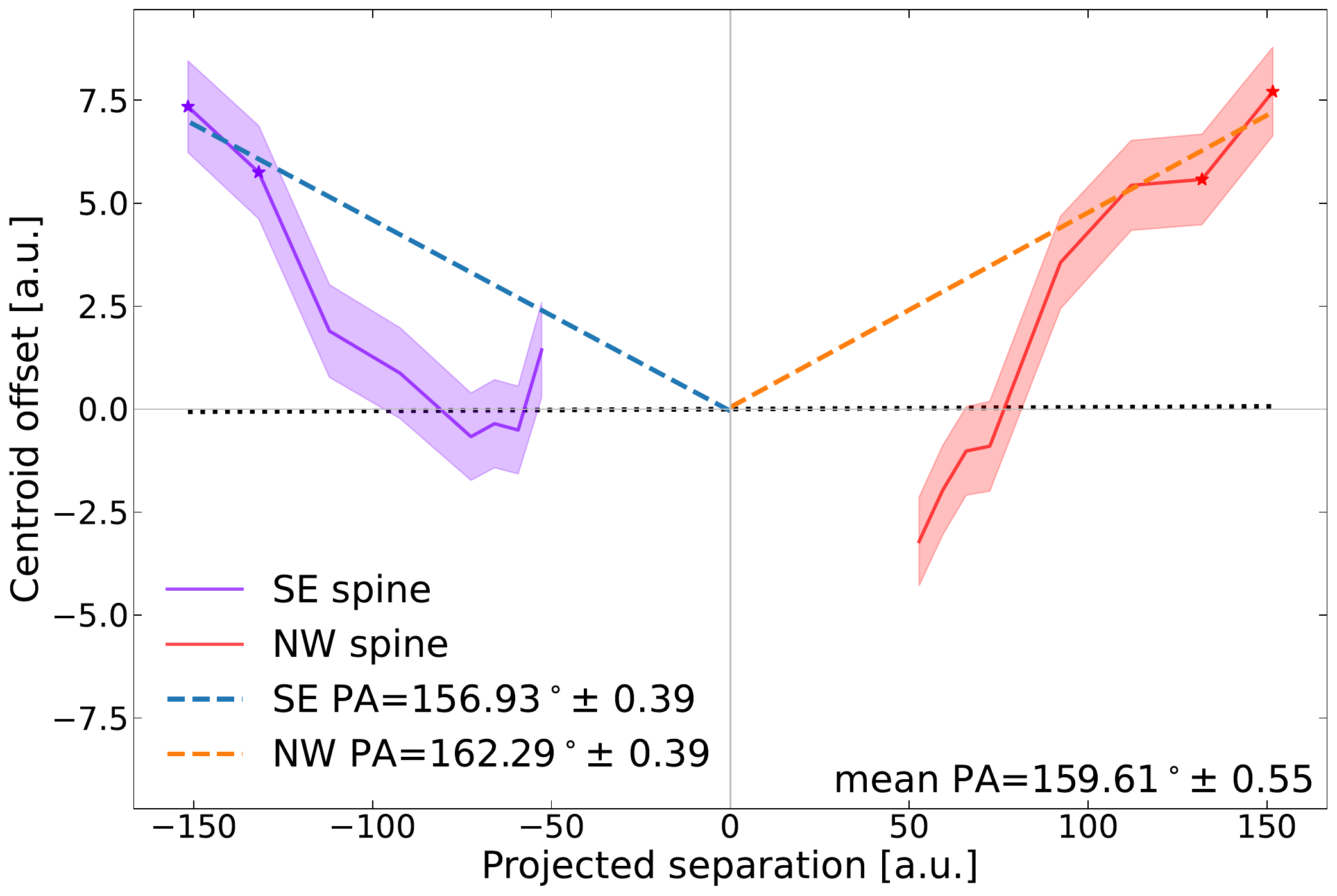}
    \caption{SE (purple) and NW (red) spine positions of the STIS observation. Star markers indicate the points used to fit the PA, which was forced through zero, and are overlaid for the SE (blue; dashed) and NW (orange; dashed) sides of the disk. Black dotted line corresponds to the mean PA.}
    \label{fig:STIS_pa}
\end{figure}

The PA of the outer disk was calculated to be $159.6^\circ \pm 0.6^\circ$, with measurements of the individual sides giving a PA of $156.9^\circ \pm 0.4^\circ$ and $162.3^\circ \pm 0.4^\circ$ for the SE and NW side respectively, as seen in Fig.~\ref{fig:STIS_pa}. The tilt angle between the two sides is $5.4^\circ \pm 0.6^\circ$, which will be discussed further in Sect.~\ref{subsec:bPic_outer}.

\begin{figure}[h]
    \centering
    \includegraphics[width=9cm]{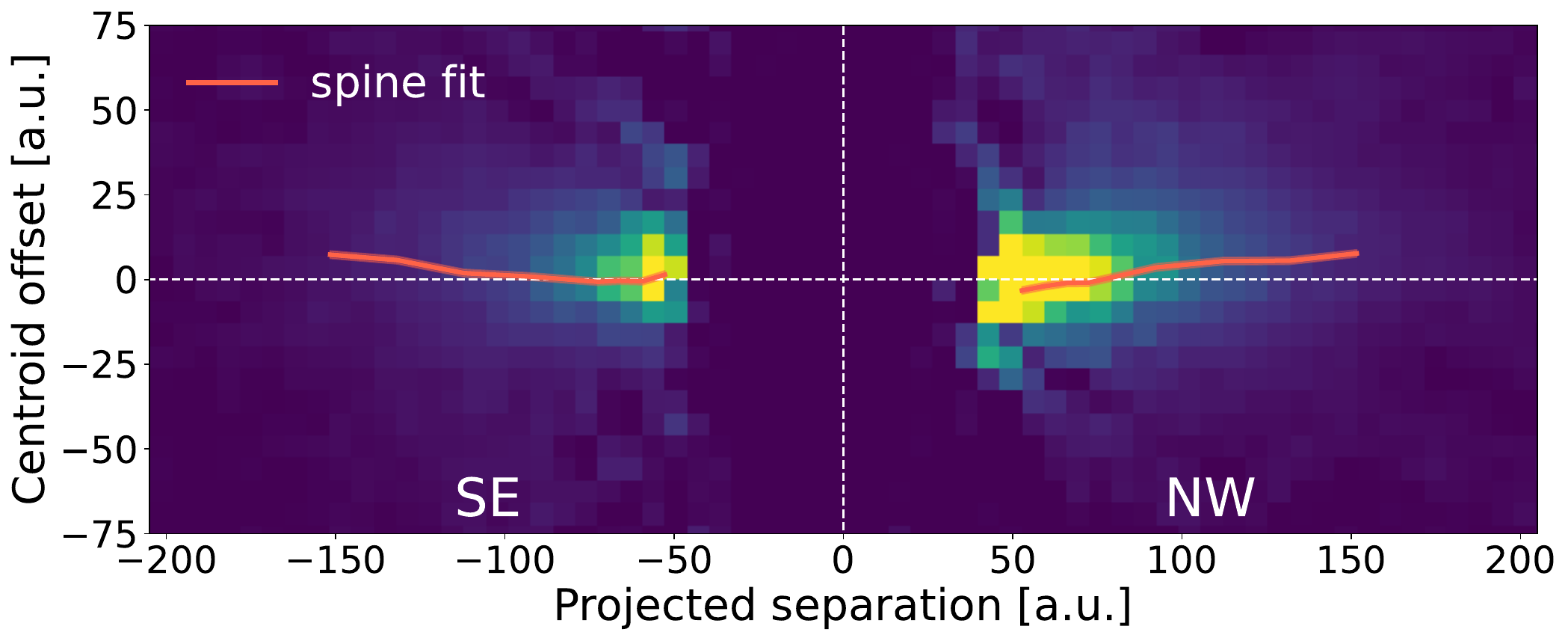}
    \caption{STIS observation of \HD{}, the image rotated to have the disk horizontal, with the position of the brightest pixels (spine) overlaid in orange. The disk PA is $159.6^\circ$.}
    \label{fig:STIS_spine}
\end{figure}

After obtaining a PA measurement, the STIS image was rotated by $70.4^\circ$ clockwise so the PA lies along the x-axis, and the spine fit repeated. The spine position, measured between 0.41$\arcsec{}$ and 1.17\arcsec{} ($53\mbox{--}152$\,au), is shown in Fig.~\ref{fig:STIS_spine}. The disk appears highly inclined with a tilt between the two sides, which is expected for a disk that is not perfectly edge-on and with anisotropic scattering \citep{Janson2021}. An asymmetry can be seen between the two sides of the disk. In the region between 0.41\arcsec{} and $\sim$\,0.6\arcsec{} ($53\mbox{--}78$\,au), the SE and NW sides lie above and below the x-axis (here passing through the centre of the star at a PA of 159.6$^\circ$) respectively, and between $\sim$\,0.6\arcsec{} and $\sim$\,0.8\arcsec{} ($78\mbox{--}104$\,au), the NW spine appears steeper than the SE spine. Beyond $\sim$\,1.0\arcsec{} (130\,au) the spine appears symmetrical on both sides.

\subsection{Morphology of the inner regions}
\label{subsec:spine}

Reductions of the SPHERE data (Figs.~\ref{fig:ADI_img} and \ref{fig:RDI_img}) show a near-edge on disk detected between $\sim$\,0.15$\arcsec{}$ and 0.6\arcsec{} ($20\mbox{--}78$\,au) with bright regions at $\sim$\,0.3\arcsec{} (40\,au) that are symmetrical on the NW and SE sides, characteristic of a ring-like structure. The disk appears to comprise of two components: a main belt between 0\,au and 40\,au corresponding to the symmetrical ring-like structure, and warped disk beyond 40\,au that extends above the plane of the belt on the NW side, and below on the SE side. This warp is clearly detected in all epochs and observations. We repeated the process used to fit the spine on the HST/STIS data in Sect.~\ref{subsec:STIS_spine} and detailed in Appendix~\ref{appendix:Gauss_fit}. The angle used to rotate the image was fixed to $70.4^\circ$, so that the inner spine could be assessed relative to the PA measured on the HST/STIS data of $159.6^{\circ} \pm 0.6^\circ$. While the disk is visible up to $\sim$\,0.6\arcsec, we only measured the spine between 0.21\arcsec{} and 0.50\arcsec{} ($27\mbox{--}65$\,au). Fitting errors in the adaptive optics (AO) system lead to a circular correction area in the image, outside of which the angular resolution is seeing limited \citep{Cantalloube2019}. The radius of this area is wavelength dependent, being in the range of $0.48\arcsec\mbox{--}0.56\arcsec$ and $0.57\arcsec\mbox{--}0.68\arcsec$ for the Y and J band wavelength channels respectively. The increased speckle brightness in this region, in conjunction with the fainter disk flux and low S/N at separations $>0.5\arcsec$, leads to difficulty in discerning whether observed structures are genuine disk signal. 

\begin{figure}[h]
    \centering
    \includegraphics[width=8.5cm]{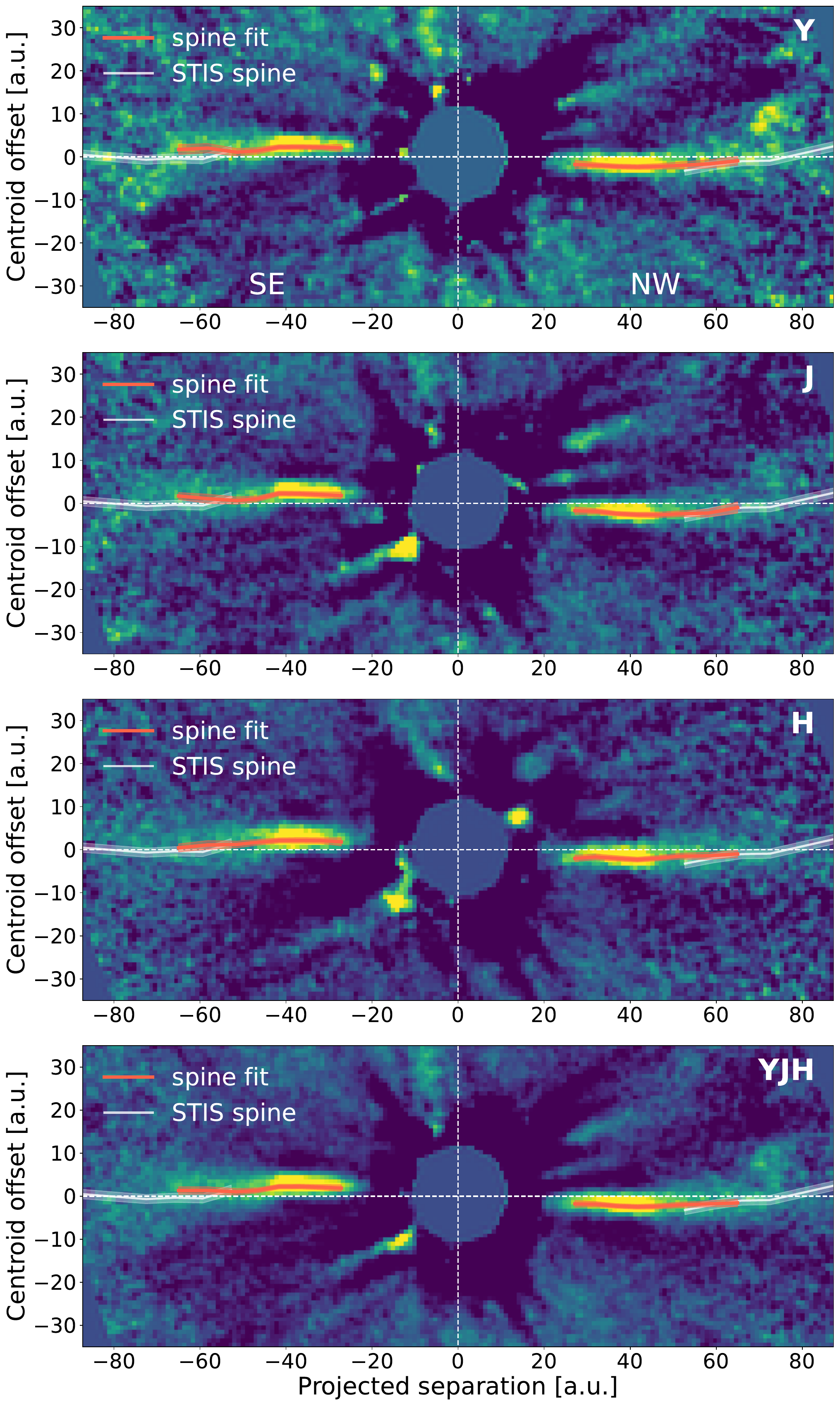}
    \caption{ADI-PCA reductions of HD110058 IFS data using 5\,PCs for the Y, J, and H band, and combined YJH data, with centroid spine position overlaid (orange). The spine measured on the STIS data in the FOV is overlaid in white. The H band image contains only the 2015-04-03 observation, while the rest combine both epochs. Images are rotated by $70.4^\circ$ clockwise.}
    \label{fig:Spine}
\end{figure}

The results of the spine fit on the Y, J, and H band, and combined YJH IFS data are shown in Fig.~\ref{fig:Spine}. After the images were rotated so that the STIS PA lies along the horizontal, we can see that the 'main belt', measured here between 27\,au and $\sim$\,40\,au, deviates asymmetrically from the mid-plane with an offset of $\sim$\,2\,au above and below the x-axis for the SE and NW sides respectively. At $\sim$\,40\,au the spine begins to tend towards the x-axis on both sides. For the NW spine in all wavebands, and SE spine in the H band, the spine continues towards the x-axis until the outer range of the fit at 65\,au, whereas in the Y and J band the SE spine tends towards the x-axis until $\sim$\,50\,au, after which it tends away from the x-axis. 

\begin{figure}[h]
    \centering
    \includegraphics[width=9cm]{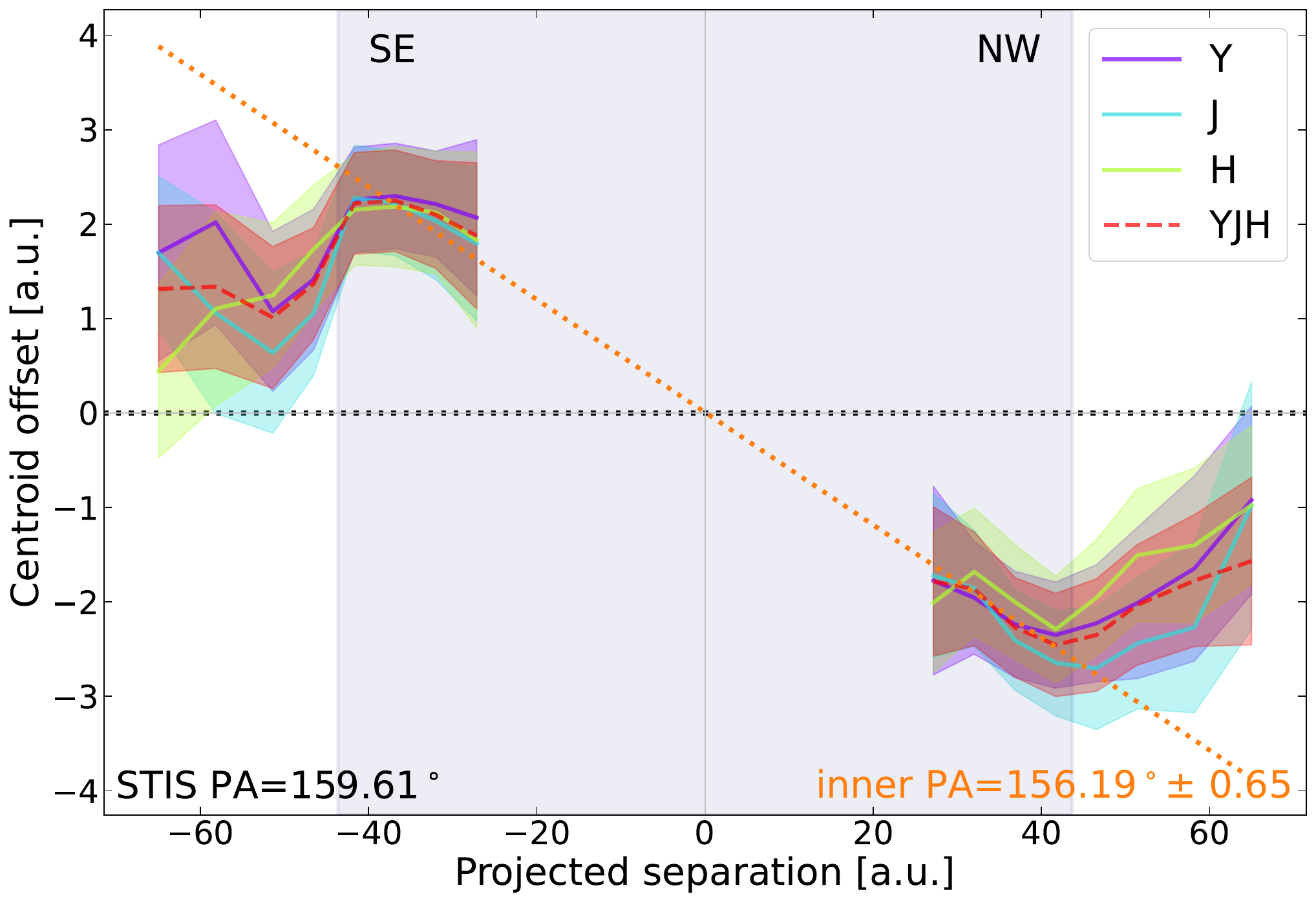}
    \caption{Spine position measured from Y (purple), J (cyan), H (lime), and combined YJH (red) band images, with the mean PA of the different bands and number of PCs fit across both sides of the disk (orange). A grey shaded box shows the region used to fit the PA of the inner disk component.}
    \label{fig:PA}
\end{figure}

The inner disk component can be fit with a continuous straight line passing through the centre of the star, albeit one that is inclined from the PA of the outer disk, from which the PA was measured. The fit was performed on the error weighted spine, as in Sect.~\ref{subsec:STIS_spine} and detailed in Appendix~\ref{appendix:Gauss_fit}, between 27\,au and 44\,au across both sides of the disk, although not forced to pass through the centre of the star. The fit was performed on the different waveband images reduced using $3\mbox{--}6$\,PCs (the range in which the disk S/N is the highest), with the final PA error calculated by combining in quadrature the measurement error, standard deviation, and true north correction error. The result of the fitting can be seen in Fig.~\ref{fig:PA}, with the PA of the inner region measured as $156.2^\circ \pm 0.7^\circ$. This results in a PA offset of $3.4^\circ \pm 0.9^\circ$ between the inner and outer parts of the disk.

\subsection{Scale height} 
\label{subsec:scale_height}

The vertical structure of debris disks offers insight into the dynamical evolution of the system \citep{Olofsson2022}. As the disk is highly inclined, it is an optimal target for measurements of its vertical structure \citep{Hughes2018}. Analysis of the scale height in scattered light complements the recent analysis of ALMA observations by \cite{Hales2022}, which traces the distribution of large dust grains and gas in the system. Their analysis found the disk to have a vertical aspect ratio of $0.13\mbox{--}0.28$ within a 99.7$\%$ confidence interval, assuming an inclination of $>80^\circ$ and the vertical aspect ratio to be constant across the disk.

We performed our analysis on the RDI reduction of the H23 data (Fig.~\ref{fig:RDI_img}) described in Sect.~\ref{subsec:RDI}. The scale height, $h$, was measured as the standard deviation of the Gaussian fit for each vertical slice, with the error in the measurement calculated using the covariance matrix of the fit. For our analysis, we quote the vertical aspect ratio, $h/r$, at the radius $r = 36.54$\,au, which is the central position of the vertical slice with the maximum $\mbox{disk flux} \times r^2$, tracing the position of maximum dust density and hence approximate radius of the planetesimal belt.

\begin{figure}[h]
\centering
    \includegraphics[width=8.5cm]{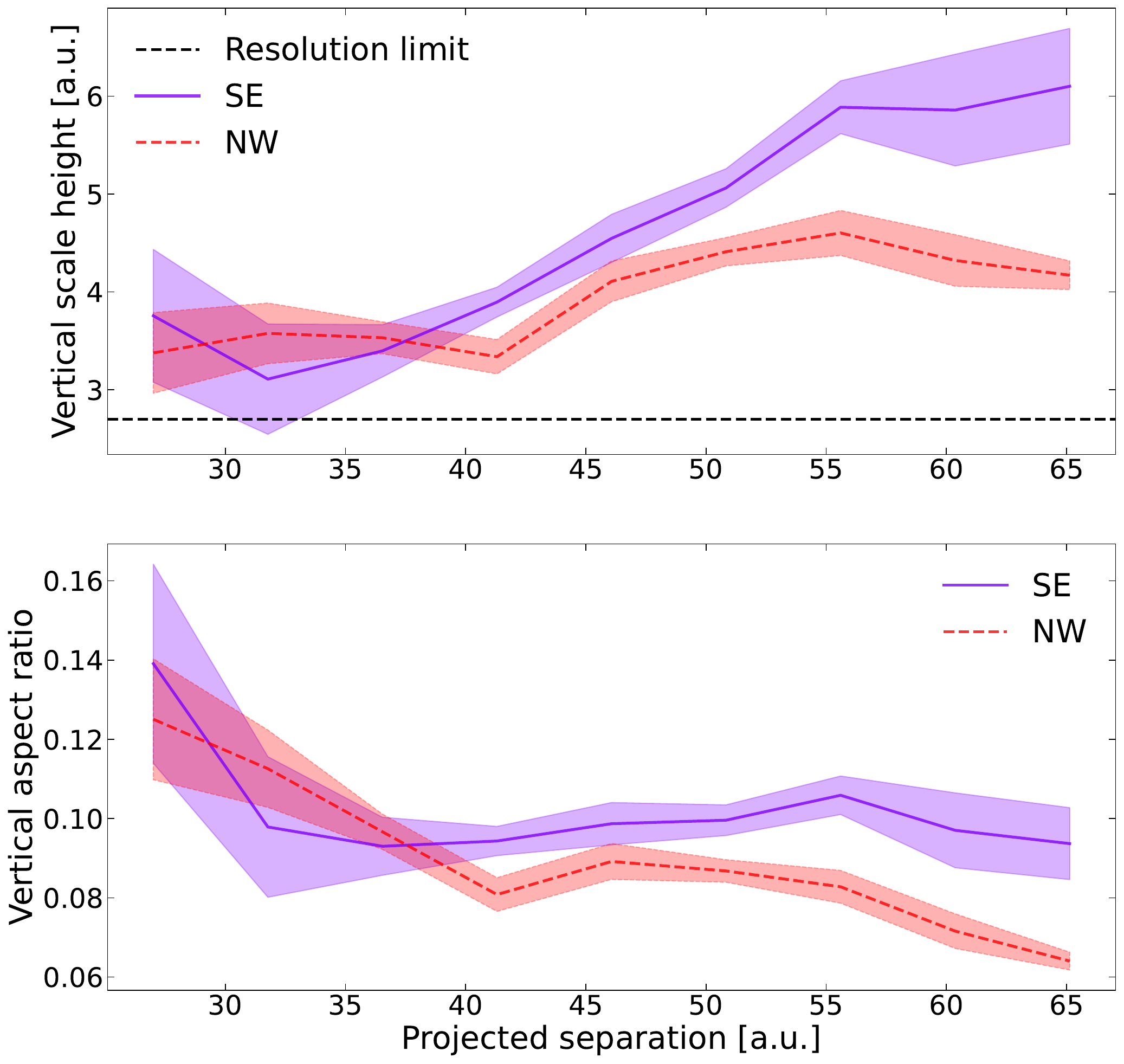}
    \caption{Disk scale height (top) and vertical aspect ratio (bottom) of RDI-PCA reduced H23 IRDIS data using 125\,PCs as a function of projected separation (SE, purple; NW, red). Scale height is expressed as the standard deviation of the Gaussian fit to a vertical slice of the disk at each separation.}
    \label{fig:vsh}
\end{figure}

The scale height and vertical aspect ratio as a function of projected separation is shown in Fig.~\ref{fig:vsh}. At $r = 36.54$\,au, the scale height was measured as $3.40 \pm 0.27$\,au ($\text{FWHM} = 8.00 \pm 0.63$\,au) and $3.53 \pm 0.16$\,au ($\text{FWHM} = 8.32 \pm 0.38$\,au) for the SE and NW sides respectively, giving a mean vertical aspect ratio of $0.095 \pm 0.009$. The FWHM of the stellar PSF is 6.24\,au for the H2 channel, and 6.46\,au for the H3 channel, suggesting the disk is marginally vertically resolved under the assumption that it is highly inclined. Interestingly, we note that the disk vertical height increases significantly beyond 40\,au on both sides. This may indicate that at this separation along the line of sight, the contribution of the warped inner disk declines in favour of the outer disk, resulting in an overall thicker vertical profile.

\subsection{Surface brightness}
\label{subsec:s_brightness}
The surface brightness along the spine of the disk was measured for both the STIS data and RDI-PCA reduced H23 IRDIS data, using the amplitude of the vertical profile Gaussian fits. For IRDIS, the amplitude was corrected for over-subtraction using fake disk analysis as described in Sect.~\ref{subsec:loss}, the flux loss ratio taken between the convolved and reduced fake disk. The corrected values were divided by the total flux of the stellar PSF and square of the plate scale to give a final measurement in contrast per square arcsecond. As the reduction of the STIS data is calibrated to physical units of Jy.arcsec$^{-2}$ and does not suffer from self- or over-subtraction of the disk, the measured amplitude of the Gaussian fit was used without further correction.

\begin{figure}[h]
    \centering
    \includegraphics[width=8.5cm]{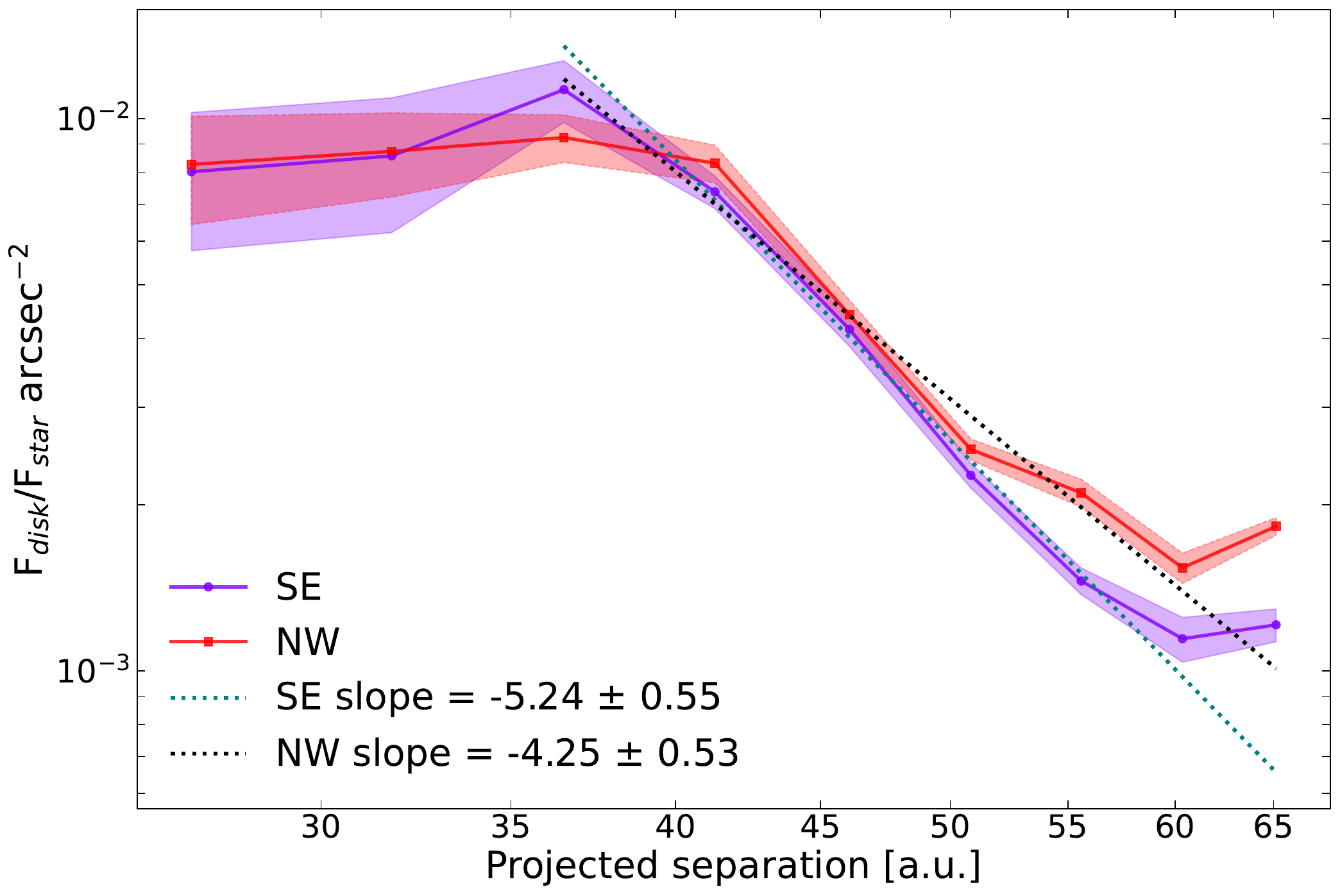}
    \caption{Surface brightness of the SE (purple) and NW (red) sides of the disk from the 2015-04-12 H23 IRDIS data reduced using RDI-PCA with temporal-mean correction and 125\,PCs. The slope of the surface brightness in logarithmic scale is overlaid with dotted lines (SE, teal; NW, black).}
    \label{fig:sb}
\end{figure}

\begin{figure}[h]
    \centering
    \includegraphics[width=8.5cm]{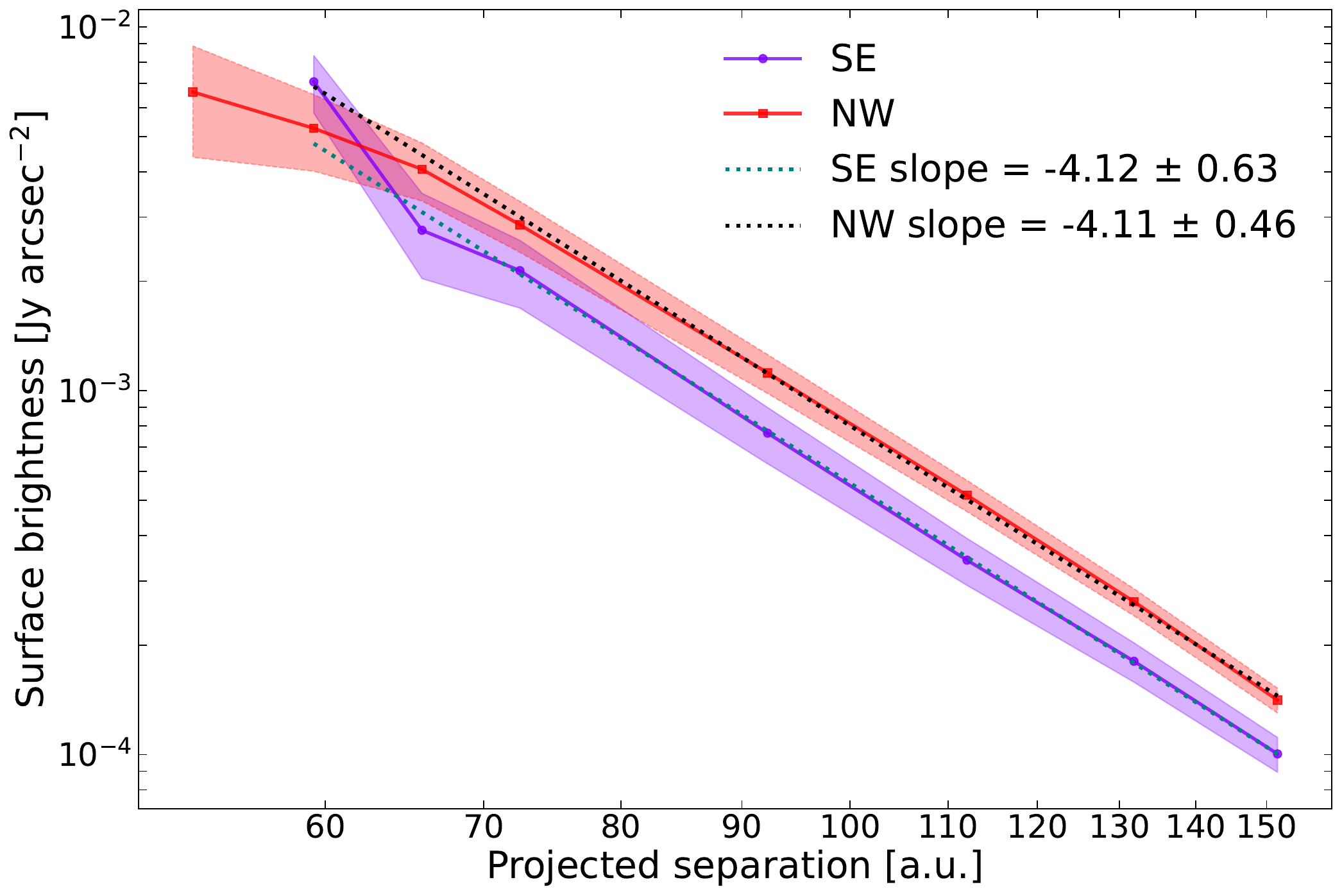}
    \caption{Surface brightness of the SE (purple) and NW (red) sides of the disk from the STIS RDI reduced data. The slope of the surface brightness in logarithmic scale is overlaid with dotted lines (SE, teal; NW, black).}
    \label{fig:STIS_sb}
\end{figure}

In Figs.~\ref{fig:sb} and \ref{fig:STIS_sb}, we show the measured surface brightness profiles of the IRDIS and STIS data. For the IRDIS data, we see an increase in the star-disk contrast starting from the smallest separation, and peaking at the bin centred at 36.5\,au on both sides, consistent with the peak radius of sub-millimetre emission \citep{Hales2022}. The SE side has a slightly higher peak contrast, which falls more quickly than that of the NW side, resulting in a smaller contrast for the SE at separations larger than $\sim$\,50\,au. For the STIS data, the surface brightness decreases with increasing separation, with a slight divergence between the SE and NW sides at separations below $\sim$\,65\,au. The slope appears steeper for the SE side, corresponding to what is seen at separations $>50$\,au in the IRDIS data. The surface brightness of the STIS data appears higher for the NW side at separations $\gtrsim$\,65\,au, which is also in agreement with the IRDIS data beyond 50\,au. We note that while we can compare the behaviour of the surface brightness as a function of separation, due to the difference in units -- flux ratio per square arcsecond for IRDIS and flux density per square arcsecond for STIS -- the measured values are not directly comparable between the two instruments.

The power-law index of the surface brightness was found using a least-squares straight line fit of the curve in logarithmic space. For IRDIS, between 41.3\,au and 65.1\,au, the power law index for the 125\,PC reduction (Fig.~\ref{fig:sb}) was measured as $-5.2 \pm 0.6$ for the SE side, and $-4.3 \pm 0.5$ for the NW side. This analysis was also repeated for reductions using 100, 150, 175, and 200\,PCs, giving a mean power-law index of $-4.9 \pm 1.2$ and $-4.2 \pm 1.1$ for the SE and NW side respectively. The measurement errors of the fits are combined in quadrature to give the final error. For each PC value tested, the SE slope in logarithmic space was steeper than that of the NW slope within 1 sigma. For the STIS data, the power-law index, between 59.3\,au and 151.6\,au, was measured to be $-4.12 \pm 0.63$ and $-4.11 \pm 0.46$ for the SE and NW sides respectively. 

These values are comparable to other edge-on debris disks in scattered light, and marginally compatible with the average surface brightness slope profile expected from theoretical predictions ($-3.5$) and numerical experiments ($-3.42$) \citep{Thebault2023} for a disk spatially resolved in the vertical direction. The possible change in the slope at $\sim$\,35\,au and $\sim$\,$60\mbox{--}65$\,au may correspond to changes in the surface density of the underlying planetesimal belt. The millimetre emission of the disk probed with ALMA favours a peak dust density at $r_c = 31^{+10}_{-8}$\,au and an outer edge $r_\text{out} = 67 \pm 4$\,au where the surface density has decreased by 50\% \citep{Hales2022}.

\subsection{Reflectance spectrum}
\label{subsec:reflectance}

We extracted the flux of both sides of the disk in each wavelength channel in the same region used for the flux loss calibration in Sect.~\ref{subsec:loss}, where the disk appears the brightest. These values were scaled by the flux loss of the fake disk injection and divided by the total stellar PSF in the respective wavebands.

\begin{figure}[h]
    \centering
    \includegraphics[width=9cm]{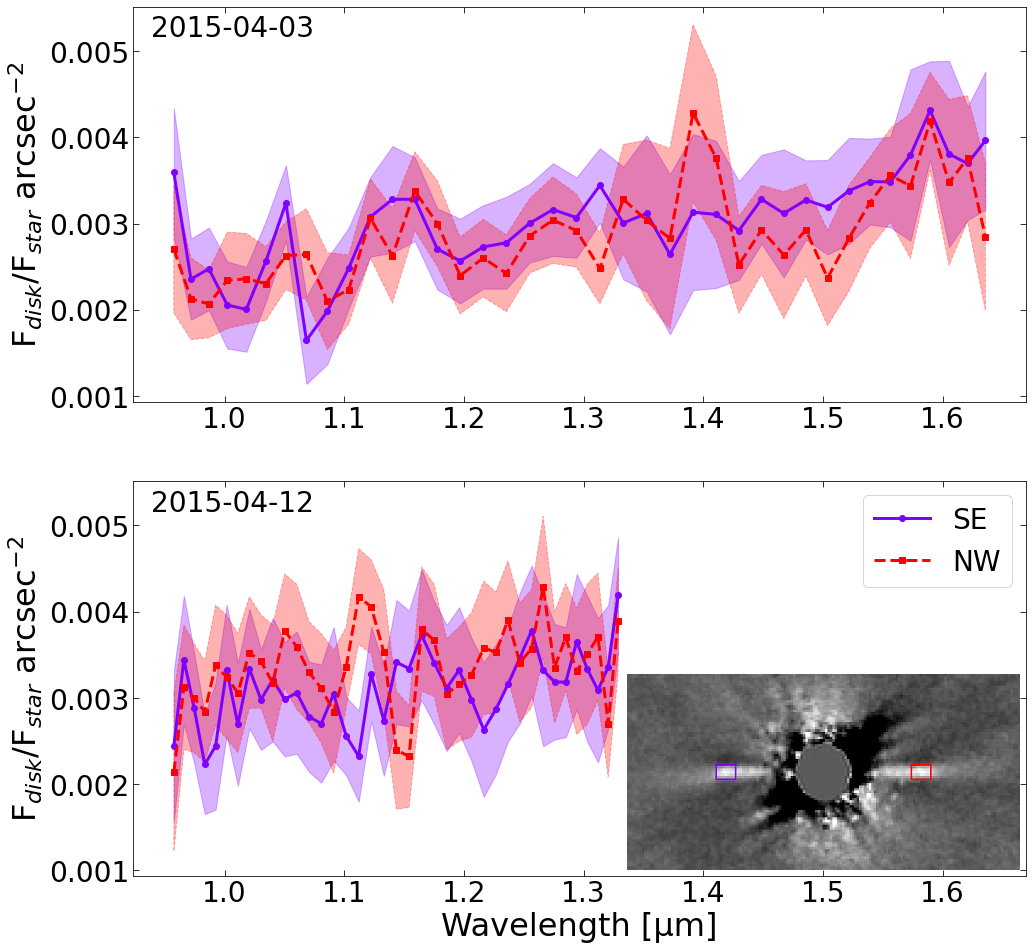}
    \caption{Disk reflectance expressed in units of (ADU.s$^-1$)$_{disk}$/(ADU.s$^-1$)$_{star}$ per arcsec$^2$ as a function of wavelength. Extracted from the 2015-04-03 (top) and 2015-04-12 (bottom) IFS data, reduced using ADI-PCA with 5 PCs and spatial-mean subtracted normalised data, for the SE (purple) and NW (red) sides of the disk, measured within the regions shown on the bottom right.}
    \label{fig:RefSpec}
\end{figure}

The disk reflectance spectrum is compatible within error bars between the SE and NW side, as shown in Fig.~\ref{fig:RefSpec}. The only $>1\upsigma$ deviation between the two sides occurs at $\sim$\,1.1\micron{} which also corresponds to a strong water vapour telluric absorption band, we therefore do not consider it significant. 
The reflectance spectrum is featureless within error bars between 1.0\micron{} and 1.6\micron{}. It is flat between 0.95\micron{} and 1.1\micron{} at a contrast of $\sim$\,0.0011\,arcsec$^{-1}$, and then linearly increasing beyond 1.1\micron{} to reach a contrast of $\sim$\,0.0018\,arcsec$^{-1}$ at 1.65\micron{}, corresponding to an increase of 48\%. The dust therefore displays a red colour in this spectral range. This red colour is compatible with the HST/NICMOS colour (F110W-F160W) of $0.2^{+0.5}_{-0.5}$\,mag derived by \citet{Ren2023}, corresponding to a $20^{+70}_{-44}\%$ flux increase between 1.1\micron{} and 1.6\micron{}. The grey colour in the 0.95\micron{} to 1.1\micron{} region is also consistent with the colour derived between the optical (HST/STIS) and 1.1\micron{} (HST/NICMOS F110W filter) within the errors, estimated as $-0.1^{+0.3}_{-0.3}$\,mag \citep{Ren2023}, and suggesting that a turning point between a blue-grey and red slope should occur at $\sim$\,1\micron{}, although we note that a fully flat spectrum is still statistically compatible with the quoted values. This specific spectral behaviour is also seen around other bright debris disks around A-type stars, for example, \bpic{} and HD\,32297 \citep{Ren2023}. 
A minimum at 1.0\micron{} followed by a red positive slope of the reflectance is a spectral feature common to many asteroid types \citep{DeMeo2009}, such as A-type asteroids typically having a steep red slope with a 100\% increase between 1.0\micron{} and 1.6\micron{}, Sa-types having a slightly shallower increase of $\sim$\,70\% and S-types with a $\sim$\,30\% increase. The minimum at 1.0\micron{} is associated with a broad olivine absorption band in A- and S-type asteroids between 0.8\micron{} and 1.2\micron{}, but we cannot confirm this feature in the \HD{} reflectance spectrum.

%
%-------------------------------------------------------------------
\section{Interpretation: Evidence for a gravitational perturbation by an inner companion}
\label{sec:interp}

The morphology described in Sect.~\ref{sec:properties} is not axisymmetric. This suggests that the central star acting on the disk via the gravitational force and the radiation pressure, two centro-symmetric forces, is not alone in shaping the disk morphology. In this section, we investigate how a gravitational perturber could explain our observed findings. 

\subsection{Comparison with $\beta$ Pictoris}

The general shape of the disk with a warp in the inner part and an offset between the two sides of the disk in the outer part is reminiscent of the nearly edge-on disk detected around \bpic{} from 50 to thousands of\,au. Here, the \HD{} disk is much more compact than \bpic{}. It is detected with STIS up to 150\,au only, with a warp detected from 30\,au to 60\,au with SPHERE. It is also noteworthy that both disks have a similar age \cite[17\,Myr and $18.5^{+2}_{-2.4}$\,Myr respectively,][]{Pecaut2012,Miret-Roig2020} and a significant amount of CO gas was detected with ALMA, likely of secondary origin \citep{Hales2022,Matra2017}. We first discuss the analogy with the outer parts of the disk and then discuss the asymmetry seen in the inner parts. 

\subsubsection{Outer regions}
\label{subsec:bPic_outer}

In early imaging data, \citet{Kalas1995} discovered an asymmetry referred to as the wing-tilt asymmetry in the outer parts of the \bpic{} disk: the spines have opposite slopes on each sides of the disk. \citet{Kalas1995} reported a $1.3^\circ$ offset between the mid-planes of the two disk extensions measured up to 20\arcsec{} (389\,au, see their Fig.~10). This asymmetry was further explored in \citet{Golimowski2006}, and more recently at larger separations thanks to a deeper data set presented in \citet{Janson2021} who show that beyond 500\,au, the two sides of the disk appear tilted by $7.2^\circ$, that is, more than at smaller distances. This tilt could be explained by a combination of the outer disk not being perfectly edge-on (inclination $i < 90^\circ$) and some anisotropy of scattering (Henyey-Greenstein scattering anisotropy parameter $g \neq 0$). These two parameters are, however, degenerate. \citet{Janson2021} find that a combination of $i = 85^\circ\mbox{--}86^\circ$ and $g = 0.8\mbox{--}0.9$ is necessary to acquire a tilt angle of $7.2^\circ$ for the \bpic{} disk.

\begin{figure}[h]
    \centering
    \includegraphics[width=9cm]{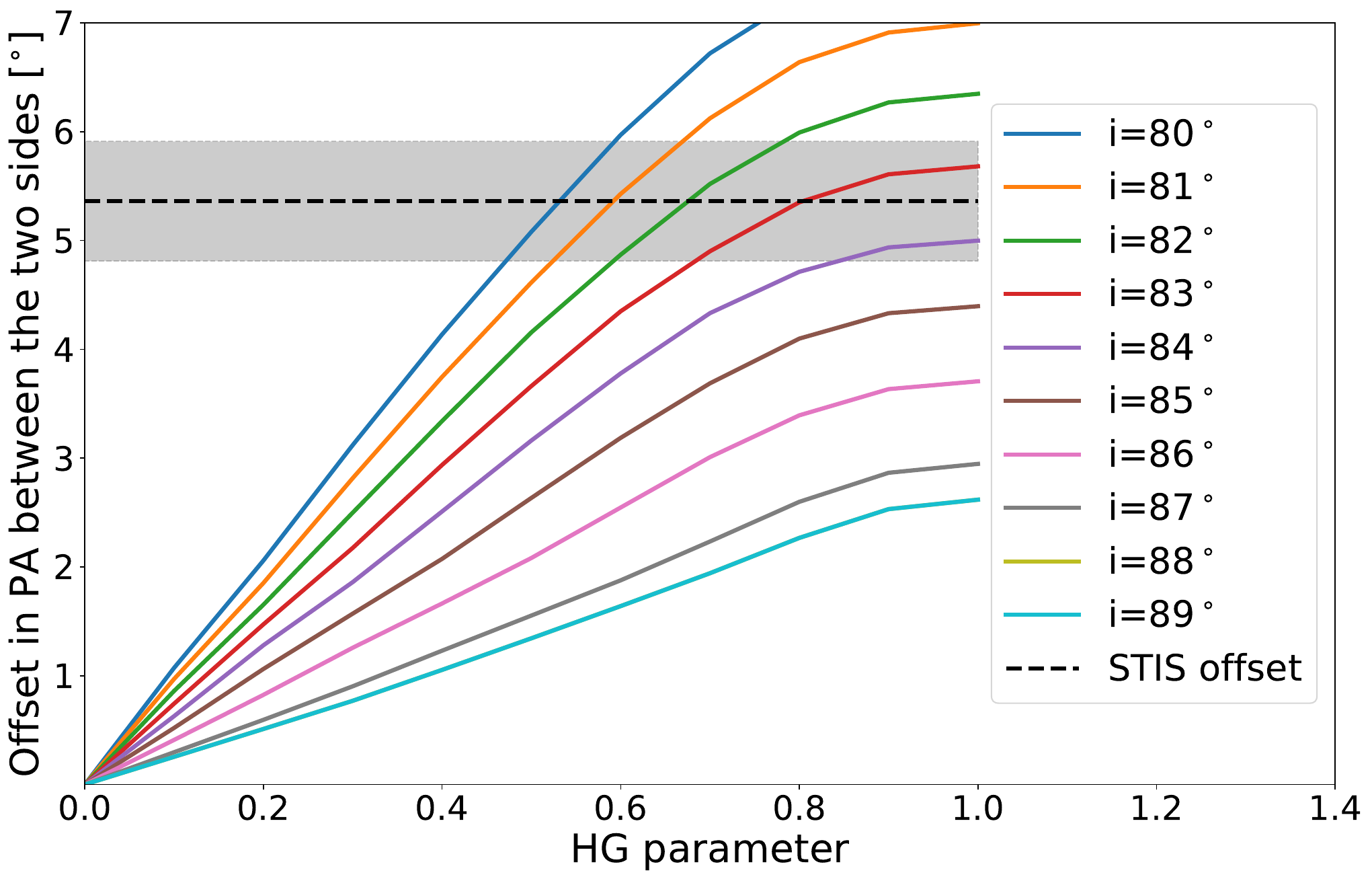}
    \caption{Offset in position angle between the two sides of a synthetic disk as a function of the Henyey-Greenstein (HG) anisotropy parameter, shown for various disk inclinations. The horizontal dotted line and the grey shaded area represents the offset measured with HST/STIS with its uncertainty.}
    \label{fig:interpretation_offset_in_PA}
\end{figure}

For the disk around \HD{}, Fig.~\ref{fig:STIS_spine} shows that the two sides of the disk are tilted by $5.3^\circ \pm 0.4^\circ$. Analysis was performed on synthetic disk models (see Sect.~\ref{subsec:loss}) to determine the values of $g$ and $i$ which are capable of reproducing such a tilt. The Henyey-Greenstein parameter was varied between 0 and 1 for disks with inclinations in the range of $80^\circ\mbox{--}89^\circ$. As with the previous analysis in Sects.~\ref{subsec:STIS_spine} and \ref{subsec:spine}, the position of the disk spine was measured using a Gaussian fit for sequential vertical profiles, and the slope of either side of the disk was measured using a 1D polynomial fit in order to derive the tilt angle between the sides. The results of this analysis can be seen in Fig.~\ref{fig:interpretation_offset_in_PA}, in addition to the tilt angle measured with HST/STIS. Reproducing the tilt angle we see with STIS requires a Henyey-Greenstein parameter $g$ in the range $0.5 < g < 1$ for an inclination $i$ in the range $80^\circ < i < 84^\circ$. This is in agreement with the best fit inclination of the ALMA continuum data derived in \cite{Hales2022} of ${78^\circ}^{+7^\circ}_{-3^\circ}$. %${78}^\circ{}^{+7}_{-3}$.

\subsubsection{Inner regions}
\label{subsec:bPic_inner}

Early ground-based \citep{Mouillet1997} and HST/STIS observations \citep{Heap2000,Apai2015} revealed the inner distortion of the \bpic{} disk in scattered light, up to about 130\,au. This warp was modelled by \citet{Mouillet1997}, \citet{Augereau2001}, and \citet{Apai2015} as being due to the result of the gravitational perturbation of the planetesimal belt by a giant planet. The inclined orbit of the giant planet forces the precession of the planetesimals’ orbit, which leads to a warp. The outer radius of the warp will increase slowly with time, as the effect of the perturbations gradually accumulates for planetesimals on larger orbits. The small dust grains generated by a collisional cascade in the warp are then subsequently blown out, creating the extended, asymmetric structure seen in detections of the disk. This giant planet was later discovered by \citet{Lagrange2009}. A later analysis \citep{Lagrange2012} showed that the $\sim$\,8~Jupiter-Mass planet's projected orbit is located above the main disk mid-plane, close to the warped component at a $3.5^\circ\mbox{--}4.6^\circ$ inclination, in agreement with the previous predictions. 

The mass-separation relationship of an inner perturber in the \bpic{} system was derived in \cite{Augereau2001} using the equation:
\begin{equation}
\label{eq:JCA}
\log \left(\frac{R_{W}}{10 \mbox{\,au}}\right) = 0.29 \log \left(\frac{M}{M_{*}} \left(\frac{D}{10 \mbox{\,au}}\right)^2\frac{t}{t_\mathrm{unit}} \right) -0.2 \qquad,
\end{equation}
where $R_{W}$, $M$, $M_{*}$, $D$, $t$, and $t_{unit}$ are the radius of the warp in the disk, the planet mass, the stellar mass, the orbital radius of the planet, the age of the system, and the time unit $\sqrt{(10 \mbox{\,au})^3/(GM_{*})} \sim$\,3.47\,years respectively. 

For the case of \HD{}, with $M_{*} = 2.1 \,\mbox{M}_{\odot}$ \citep{Chen2014}, and $t = 17$\,Myr, we find $MD^2 = 39.34\,\mbox{M}_{\mathrm{Jup}}\,\mbox{au}^2$ for a warp radius $R_{W} = 45$\,au. A \bpic{}\,b analogue of $8\,\mbox{M}_\mathrm{Jup}$ would need to have an orbital radius of 2.2\,au (0.02\arcsec) to produce the warp we see in \HD{}. This is in agreement with the secular perturbation analysis performed by \cite{Hales2022} using an alternate equation to Eq.~\ref{eq:JCA}.

There are currently 35 confirmed companions with semi-major axes between 1.5\,au and 3.0\,au that have masses in the range of $6\mbox{--}12\,\mbox{M}_\mathrm{Jup}$, the vast majority detected via the radial velocity method \cite[NASA Exoplanet Archive,][]{ExoArchive}. With \HD{} being 6.5 times more distant than \bpic{}, we cannot probe the regions inside 20\,au (0.15\arcsec) with SPHERE to search for a possible giant planet (see Sect.~\ref{sec:planets} for the detection limits). Such a scenario will be discussed through dynamical simulations in Sect.~\ref{subsec:dynamical_model}.

\subsection{Spine position using HST/STIS and VLT/SPHERE}
\label{subsec:combined_spine}

Observations with HST/STIS and VLT/SPHERE are very complementary to each other, as the former is sensitive to the halo of small bound grains at large distances from the central star, while the latter probes the innermost regions of the disk, including the birth ring and possibly some of the halo. While different dust size populations are imaged in the optical than in the NIR, our measurements of the spine (Figs.~\ref{fig:STIS_spine} and \ref{fig:Spine}) likely trace where the dust is most congregated, and so are comparable for the different instruments. 

\begin{figure}[h]
    \centering
    \includegraphics[width=9cm]{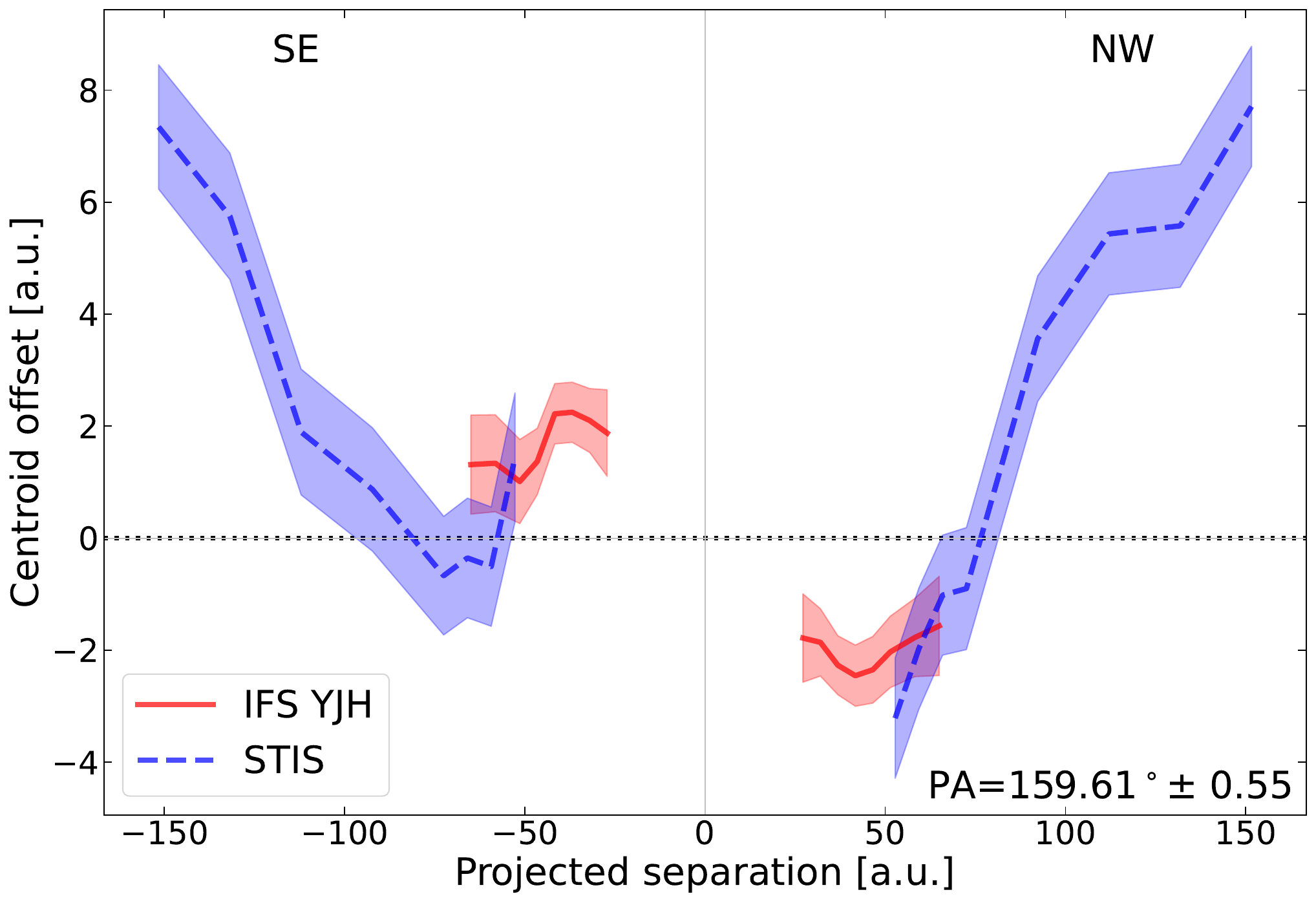}
    \caption{Spine position of HST/STIS optical data (blue; Fig.~\ref{fig:STIS_spine}) and IFS YJH band data (red; Fig.~\ref{fig:PA}) derotated by $70.4^\circ$ so that the PA lies along the horizontal.}
    \label{fig:IFS_STIS_spine}
\end{figure}

Figure~\ref{fig:IFS_STIS_spine} shows the spine measured on both the STIS and IFS data sets. Due to the low S/N of the disk at larger separations in the NIR data restricting the region of the spine fit to $<70$\,au, the fitting only overlaps between 53\,au and 65\,au for the two instruments. The NW spine measured on the STIS and IFS data in this region are in good agreement, while the SE spine is not. This is unsurprising; the spine fitting on the SPHERE data deviates between the different wavebands most prominently on the SE side beyond $\sim$\,50\,au (Fig.~\ref{fig:Spine}), which suggests the fit is being affected by wavelength dependent noise obscuring the disk flux. 

From our analysis of the wing-tilt of the outer disk in Sect.~\ref{subsec:bPic_outer}, we expect the disk to have an inclination $i < 84^\circ$; however, the inner component does not show the bowed shape we would expect for a disk with anisotropic scattering at that inclination \citep{Milli2014}. The region of the disk we expect to be bowed is within 0.3\arcsec, which is more affected by ADI reduction artefacts that serve to artificially distort the shape of the disk than regions at larger separations, and could cause this feature to be lost or flattened. Conversely, this could also indicate that the warped component of the disk is at a different inclination from the outer disk. It is likely that the argument of periapsis of a disk will not be in the plane of the sky, and so a tilted inner component would appear as a disk with a different inclination to the outer component in addition to the PA offset. Furthermore, an unresolved observation of a two-component disk with different inclinations could offer an additional explanation to the large scale height found with ALMA, discussed in the following Sect.~\ref{subsec:scale_height_ALMA}.

\subsection{Scale height in thermal and scattered light}
\label{subsec:scale_height_ALMA}

Both \HD{} and \bpic{} have been imaged in the continuum emission with ALMA, for the latter, \cite{Matra2019} fit the disk of \bpic{} using several models all of which agree upon a relatively small vertical aspect ratio $h = 0.07 \pm 0.02$. For \HD{} on the other hand, \cite{Hales2022} and \cite{Terrill2023} find a relatively large vertical aspect ratio of $h = 0.17^{+0.05}_{-0.09}$ and $h = 0.21 \pm 0.03$, respectively. This is in contrast with the scale height of the disks in scattered light, for which both appear vertically thin. 

A likely scenario for \HD{} appearing vertically thick at millimetre wavelengths while remaining thin in scattered light is proposed by \citet{Hales2022} and requires the gas density to be high enough, so that small dust with Stokes numbers close to 1 can migrate out and also sediment due to gas-drag \citep{Olofsson2022}. The gas density should however not be too high otherwise the Stokes number becomes $\ll$\,1 and small dust stay coupled to the gas as in protoplanetary disks. The CO observations are compatible with the former scenario, with CO gas emission ranging mostly from $\sim$\,$1\mbox{--}10$\,au out to 30\,au, with a peak at $10\mbox{--}20$\,au while the dust peak radius is $31 \pm 10$\,au. In addition, if CO dominates over carbon and oxygen, \citet{Hales2022} show that the gas density is expected to be in the range $2\times10^{-4}\mbox{--} 7\times10^{-2}$\,g.cm$^{-2}$, a range where the smallest dust grains will have Stokes numbers close to 1 and thus will tend to settle towards the mid-plane, producing a vertically thin disk in scattered light.
Interestingly, the models proposed by \citet{Hales2022} to reproduce the spatially segregated distribution of CO and continuum emission in the disk based on secondary CO gas require either a low gas viscosity smaller than $10^{-4}$, or a massive planet in the inner 10\,au accreting most of the in-flowing gas, which could be the same as that also responsible for the warp detected in our data.

\subsection{Dynamical models supporting an inner perturber} 
\label{subsec:dynamical_model}

Here we carry out a dynamical analysis in order to shed light on the origin of the warp in this debris disk. More specifically, we aim to set constraints on the dynamical perturber at play, and possibly make predictions as to where and how this perturber should be searched for.

\begin{figure*}
    \centering
    \includegraphics[width=0.44\linewidth]{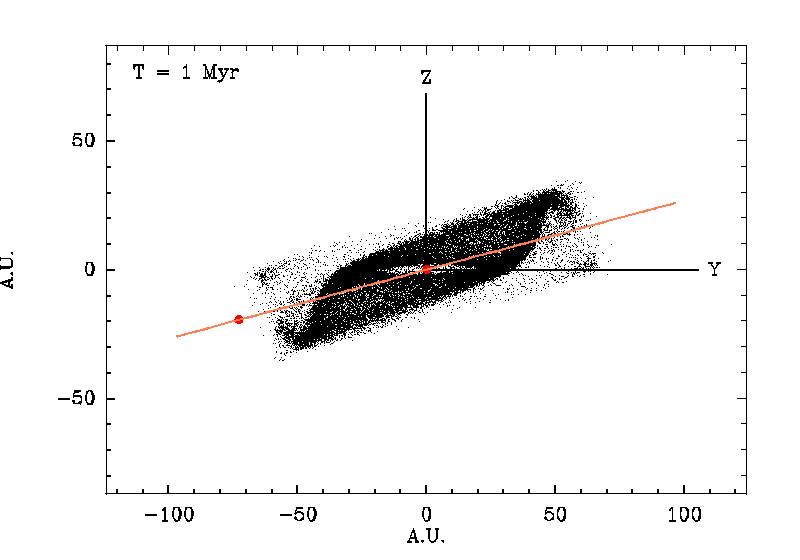}
    \includegraphics[width=0.44\linewidth]{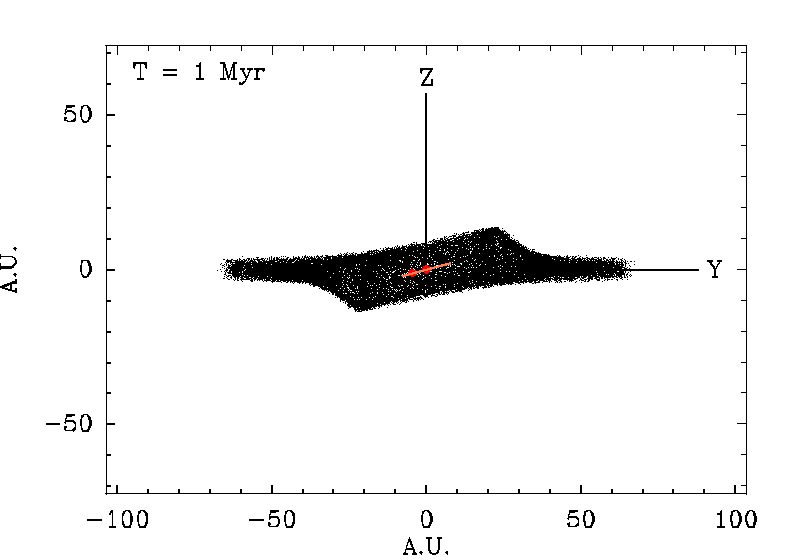}

    \includegraphics[width=0.44\linewidth]{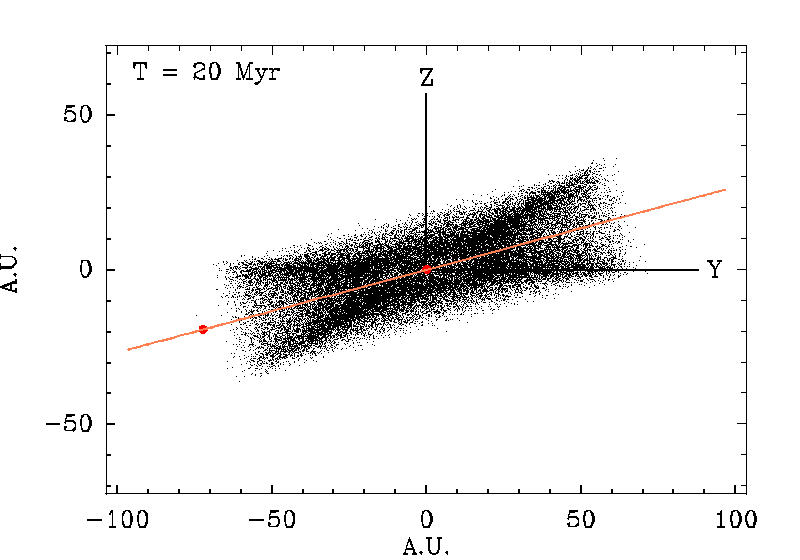}
    \includegraphics[width=0.44\linewidth]{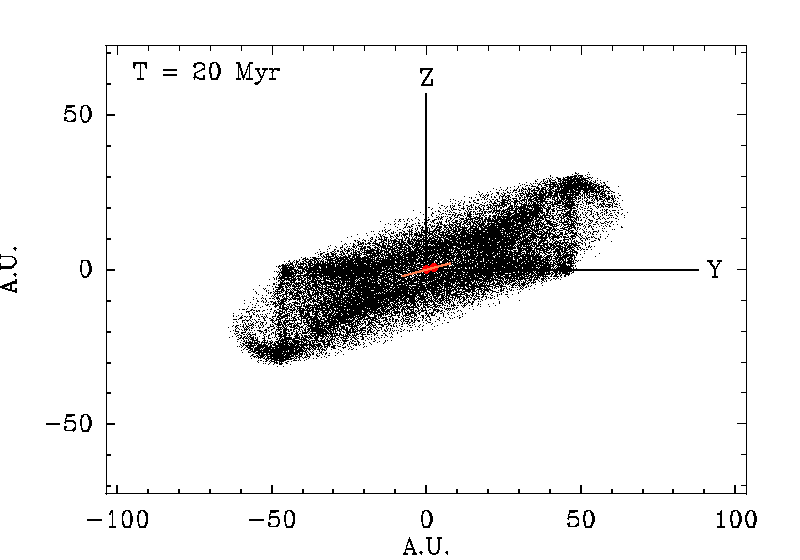}    

    \includegraphics[width=0.44\linewidth]{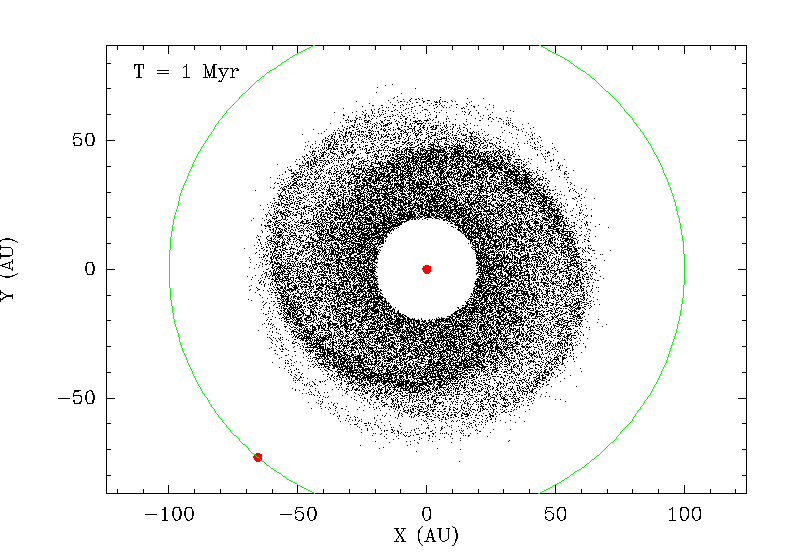}
    \includegraphics[width=0.44\linewidth]{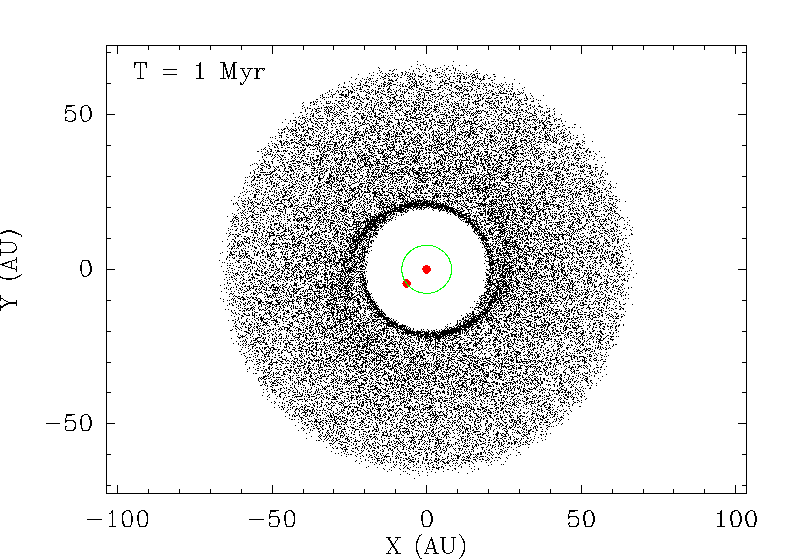}

    \includegraphics[width=0.44\linewidth]{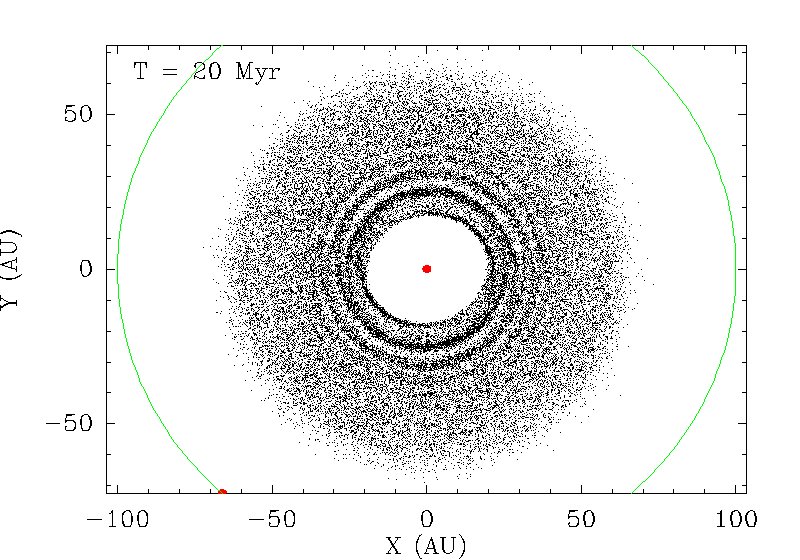}
    \includegraphics[width=0.44\linewidth]{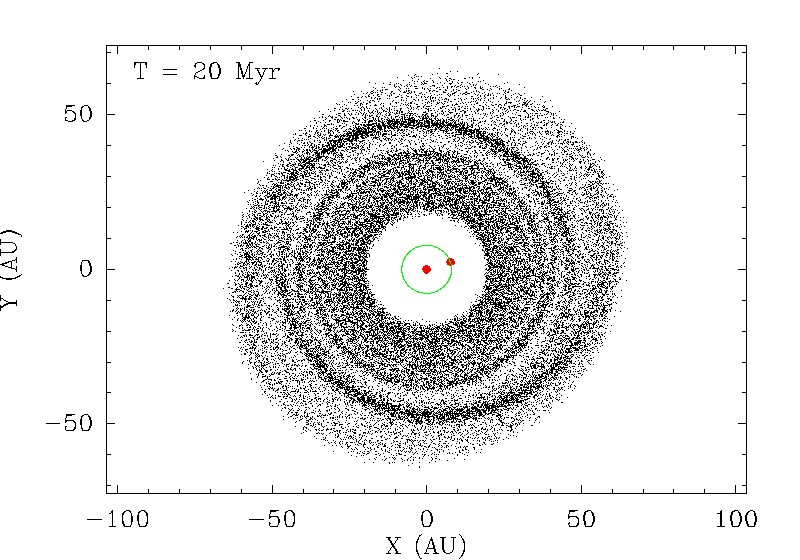}    
    \caption{Onset of warp in a disk in the case of a perturber outer to the disk (left panels) and inner to the disk (right panels); orange and green lines trace the orbit of the perturber. Systems at both 1\,Myr and 20\,Myr are shown as labelled in the panels. This onset is accompanied by spiral patterns that are not necessarily visible in the case where the disk is oriented (close to) edge-on (see upper panels) but are more readily visible in pole-on views (see bottom panels). In the case of an outer perturber, the spiral pattern and the onset of the warp evolve from the outer parts of the disk towards the inner parts, and conversely in the case of an inner perturber.}
    \label{fig:Warp_Setting}
\end{figure*}

In Fig.~\ref{fig:Warp_Setting}, we display the results of two dynamical simulations carried out using the symplectic code SWIFT-RMVS \citep{Levison1994}. In both cases, we used a central star of mass $2.1\,\mbox{M}_\odot$, surrounded by a set of 50\,000 massless test (non-interacting) particles, with semi-major axes randomly and uniformly distributed between 20\,au and 65\,au. Their initial inclinations and eccentricities were uniformly distributed between $\pm3^\circ$ and $0\mbox{--}0.05$, respectively. This mimics an initially cold disk with low inclinations and eccentricities as can be expected at the end of the protoplanetary phase, during which the action of the gas on large (mm-size and above) solid particles leads to circularisation of orbits and settling of solids in the mid-plane. 
As for the remainder of their orbital elements -- longitude of periastron $\omega$, longitude of ascending node $\Omega$, and mean anomaly $M$ -- they were randomly and uniformly distributed between 0 and $2\pi$. 
The simulations each included a planetary perturber mutually inclined with the disk by $15^\circ$. In one case, this perturber was $1\,\mbox{M}_\mathrm{Jup}$ and inner to the disk -- with a semi-major axis of 8\,au (62\,mas in projected separation, corresponding to a warp radius of 60\,au after 17\,Myr using Eq.~\ref{eq:JCA}), and in the other case, the perturber was $1.5\,\mbox{M}_\mathrm{Jup}$ and outer to the disk -- with a semi-major axis of 100~au. 
Both simulations were integrated over 20\,Myr, with snapshots taken every 1\,Myr. The timestep used was 1/20th of the smallest orbital period involved in each case.

Similarly to the system of \bpic{}, the warp morphology in the disk surrounding \HD{} hints at the dynamical shaping by a massive perturber orbiting inside the disk, and in which the warp onset propagates from the inner parts of the disk to the outer parts, as illustrated in Fig.~\ref{fig:Warp_Setting}. 

This assessment can also be understood using the Laplace-Lagrange theory, that describes the secular dynamical interactions between a massive perturber and much less massive objects, such as the kilometre-sized parent planetesimals that underlie the dust population visible with facilities and instruments such as SPHERE or ALMA. It is in this analytical frame that one can understand how an eccentric or an inclined perturber can force a whole disk's eccentricity or inclination, respectively \citep[see, e.g.,][]{Murray1999}. 
The characteristic timescale $t_\mathrm{sec}$ for the onset of these secular patterns -- that is, in the case of an inclined perturber, for the warp to propagate through the whole disk -- inversely depends on both the mass of the perturber, and on the ratio $\alpha$ between the perturber's semi-major axis $a_p$ and that of the small body it acts upon, $a$ \citep[see][]{Brady2023}: 

\begin{equation}\label{eq:tsec}
    t_\mathrm{sec} \approx \frac{4}{3} T_p \left( \frac{M_p}{M_{*}} \right) ^{-1} \alpha ^{-7/2} \qquad,
\end{equation}

\noindent where $T_p$ is the perturber's orbital period (in years), $M_p$ and $M_{*}$ are the masses of the pertubers and the star, respectively. For the disk's constituents to develop the orbital inclination imposed by the pertuber at a given distance $a$, it will take a few $t_{sec}$.
We note that in the case where the perturber orbits inside of the small body, this ratio is $\alpha = a/a_p$, and conversely, $\alpha = a_p/a$ if the perturber orbits outside of the small body, such that $\alpha < 1$, always.

This last assertion is particularly important to understand the onset location and direction of propagation of a warp throughout a debris disk. Indeed, it means that small bodies that are the closest to the perturber will be influenced the fastest. In other words, in the case of an inner perturber, the warp will start to form in the inner parts of the disk and propagate outwards, and conversely for an outer perturber, the warp will start from the outer parts of the disk and propagate inwards, as illustrated in Fig.~\ref{fig:Warp_Setting}. 

\begin{figure}
    \centering
    \includegraphics[width=0.9\linewidth]{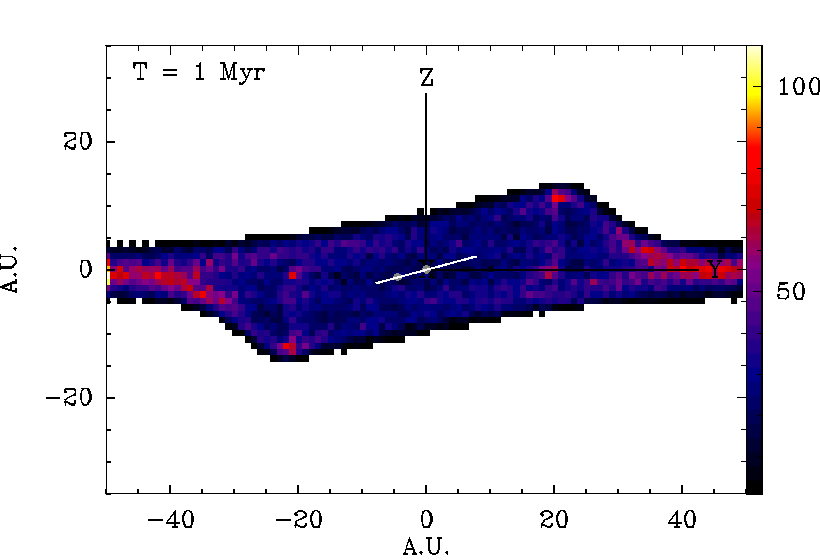}

    \includegraphics[width=0.9\linewidth]{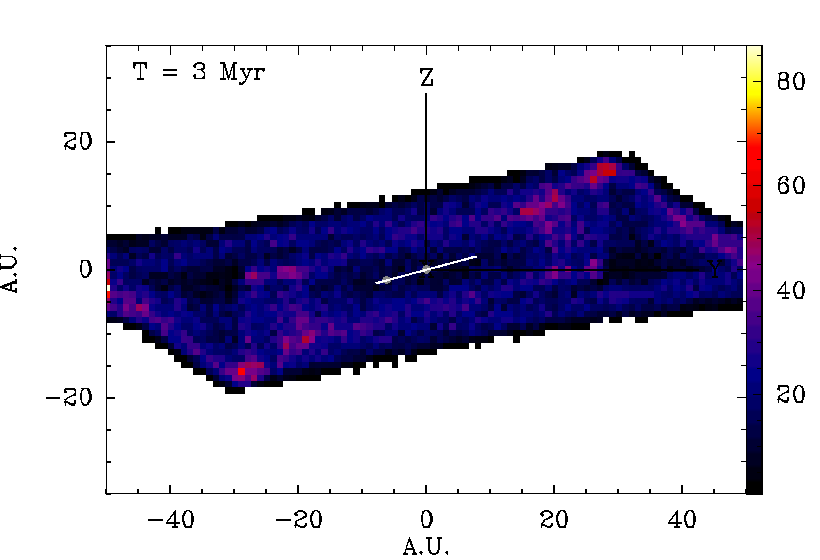}    

    \includegraphics[width=0.9\linewidth]{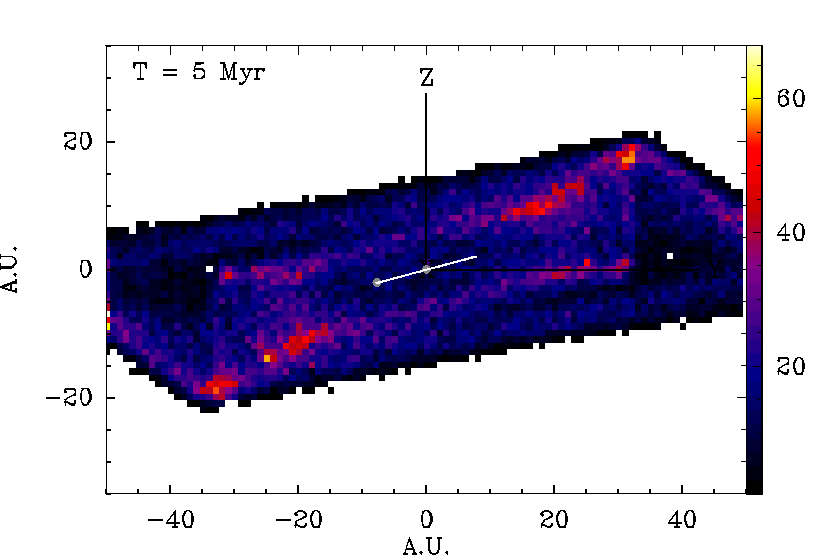}

    \includegraphics[width=0.9\linewidth]{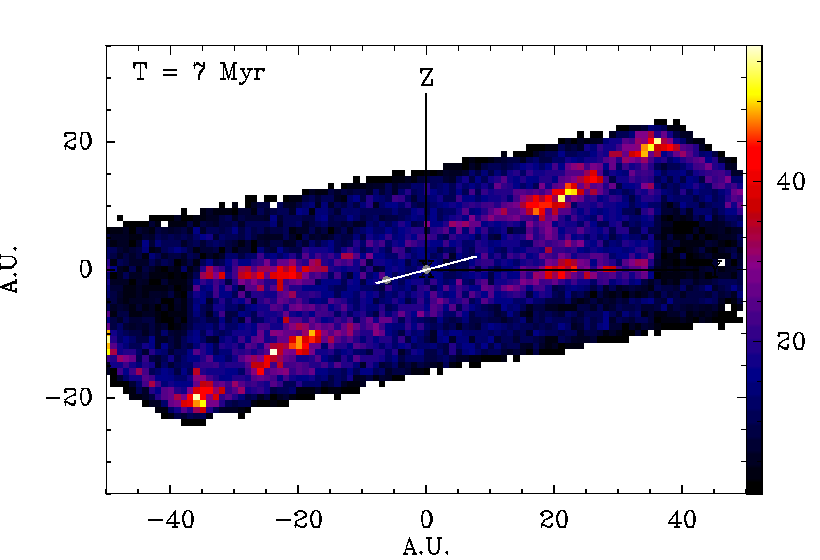}
    
    \caption{Density maps extracted from our N-body simulation with a perturber inner to the disk, its orbit shown by the white line. We oriented the disk edge-on and show snapshots at 1, 3, 5, and 7\,Myr. The warp visibly extends outwards progressively and its location is $\sim$\,20\,au, $\sim$\,30\,au, $\sim$\,35\,au, and $\sim$\,38\,au, respectively.}
    \label{fig:Warp_Evol}
\end{figure}

Naturally, and as illustrated in this figure, the spiral patterns are better visualised in pole-on views. We note that these spirals are not necessarily associated uniquely with warps. In other words, the detection of spirals in a debris disk seen pole-on does not guarantee that the planet creating them is on an orbit mutually inclined with the disk. Assertion of this orbital parameter for planets associated with debris disks through their gravitational patterns can only be made through the detection of a warp, which is best seen in a disk seen edge-on. It is also in this configuration that one can clearly see the warp evolution and progressive extension, as illustrated in Fig.~\ref{fig:Warp_Evol}. In this set of density maps extracted from our simulation using a planetary perturber inner to the disk, one can see the warp propagating outwards progressively. These maps can be used to test Eq.~\ref{eq:JCA}, the mass-separation relationship of an inner perturber derived by \citet{Augereau2001}. 

The snapshots have been taken at 1, 3, 5, and 7\,Myr, and exhibit a warp extending up to $\sim$\,20\,au, $\sim$\,30\,au, $\sim$\,35\,au, and $\sim$\,38\,au, respectively. The left hand side of Eq.~\ref{eq:JCA} then yields $\log \left(R_{W}/10 \mbox{\,au}\right)=$ 0.30, 0.48, 0.54, and 0.58, when $R_W$ is substituted with each of those values, respectively.
On the right hand side of the equation, the mass of the planet $M$, the stellar mass $M_{*}$, and the distance $D$, can be substituted with $1\,\mbox{M}_\mathrm{Jup}$, $2.1\,\mbox{M}_\odot$, and 8\,au, respectively. As for the time $t$, we use the time of the snapshot (in years) -- the time unit being 5.2\,years. This yields the following values for the right hand side: 0.31, 0.45, 0.51, and 0.55. Considering the unavoidable uncertainties pertaining to measuring the warp location on a density map by hand, these values can definitely be considered in accordance with the equation derived in \citet{Augereau2001}.

As shown in Fig.~\ref{fig:IFS_STIS_spine}, the disk spine is not axisymmetric in the inner 60\,au while the planetesimal belt extends beyond this radius as probed with ALMA. This suggests that a warp has not been fully onset throughout the whole radial extent of the disk. In other words, this debris disk has not reached a steady state yet, otherwise, it would be completely tilted \citep{Brady2023}. 

We note that our simulations do not include radiation pressure or collisions, and consequently, do not trace the smaller grains seen in the scattered light images. This means that while they are useful to show what type of pattern is generally expected in a disk gravitationally perturbed by an inclined planet -- and how this pattern evolves -- they cannot be used to provide a more direct comparison with the SPHERE observations (i.e. to generate synthetic SPHERE observations that can be directly compared with the actual observations; in particular, the production of residuals). Therefore, while helpful in the broad interpretation of the warp, features visible in the simulations are not necessarily expected to be seen in the SPHERE observations. We also note that we might over-predict the strength of some substructures, since collisions are not accounted for. That being said, we can use the results of these simulations as a proxy to compare them with the ALMA observations. In particular, \citet{Brady2023} predict that debris disks under the influence of planets significantly inclined with the disk ($\gtrsim$\,$10^\circ$) will exhibit a puffy appearance and unusual scale height when seen at ALMA wavelengths. This is in accordance with the disk morphology seen with ALMA reported by \citet{Hales2022}, and hence also supports the presence of a planetary perturber on an orbit significantly inclined compared to the disk.

Finally, as mentioned in Appendix~\ref{appendix:flyby}, two stars have been identified which approached \HD{} at distances shorter than 2\,000\,au less than 100\,000\,years ago. We can use Eq.~\ref{eq:tsec} to evaluate the timescales involved in the onset of a warp induced by either one of these closest approaches, and compare those with their actual time of closest approach. This allows us to evaluate the possibility that the warp was created by one of them. Semi-major axes $a_p$ of 856\,au and 1691\,au translate into orbital periods $T_p$ of 36\,293\,years and 100\,769\,years, respectively.\footnote{Using Kepler's third law and considering the fact that the central star is $2.1\,\mbox{M}_\odot$, one immediately derives $T_p^2=2.1 \times a_p^3$.} Assuming a mass of $1\,\mbox{M}_\odot$ for both perturbers, and evaluating the secular timescale at the outermost part of the disk ($a=70\,$au), Eq.~\ref{eq:tsec} eventually yields values of $\sim$\,310\,Myr and $\sim$\,9\,Gyr, respectively. These values are orders of magnitude larger than both the time of closest approach and the age of the system. Hence, even if the flybys approached with an inclination that could match that of the warp, they could not create it.

It is clear from both the SPHERE and ALMA observations that the debris disk surrounding \HD{} harbours a warp, and has thus been dynamically perturbed by a massive body on a mutually inclined orbit with the disk. While theoretical considerations show that a flyby-induced origin for this warp can be discarded, without a clearer picture of these dynamical imprints, it remains difficult to set constraints on the perturber at play. Extensive dynamical modelling that goes beyond the scope of this paper and higher-resolution ALMA observations would help in this endeavour.

A large mutual inclination between a debris disk and a planet, however, is not unlikely in planetary systems. Recently in the HD\,114082 system, an 8\,M$_\text{Jup}$ giant planet was found transiting the star, while the debris disk resolved in this system is inclined by $7^\circ \pm 1.5^\circ$ from edge-on \citep{Engler2022}.

%
%---------------------------------------------------------
\section{Constraints on planetary mass companions}
\label{sec:planets}
%Detection limits + conversion to mass. 

We detect three point sources at large separation beyond 2.5\arcsec{} in the IRDIS field of view. We detail their astrometry and contrast in Appendix~\ref{appendix:astrometry}. Their position in a colour-magnitude diagram shows they are likely background objects. Point source 3 in Table~\ref{tab:astrometry}, at a separation of $\sim$\,6.2\arcsec, is detected by \cite{GaiaDR3} with a G-band magnitude of $19.102 \pm 0.002$; however, no parallax or proper motion data are available. 

To bring the tightest constraints on the presence of faint companions in the scattered light images of the system, the SPHERE IRDIS and IFS data obtained in the two epochs were also reduced with the more aggressive algorithm PACO ASDI \citep{flasseur2020_PACO_ASDI}, as described in \citet{Chomez2023}. The $5\upsigma$ contrast maps are directly produced by the PACO algorithm. The contrast was converted to mass using the COND evolutionary models \citep{Baraffe2003}, assuming an age of 17\,Myr and stellar magnitudes of 7.643, 7.587, and 7.583 for the J, H, and K filters, respectively \citep{Cutri2003}. Following the guidelines detailed in \citet{Vigan2015} to derive mass detection limits from multi-spectral data, we converted our IFS-YJH contrast using the stellar magnitude in the broad-band H filter, and our IFS-YJ contrast using the stellar magnitude in the broad-band J filter.
 
To further constrain the presence of companions in the system, we combined the SPHERE direct imaging sensitivity maps with GAIA astrometry. We used the Multi-epochs multi-purposes Exoplanet Simulation System 3 (MESS3) tool (Kiefer~et~al. in prep.), an extension of the MESS2 tool \citep{Lannier2017} that combines direct imaging and radial velocity data. The new MESS3 release allows us to use the excess noise and proper motion anomaly from GAIA DR3 \citep{GaiaDR3} to constrain the presence of bound companions. The proper motion anomaly is described, for instance, in \citet{Kervella2019,Kervella2022}. The excess noise represents the difference between the dispersion of the estimated Gaia measurements at the estimated times of observations and the instrumental/intrinsic Gaia noise (Kiefer~et~al. in prep). We used MESS3 to constrain the presence of companions in the system HD\,114082 \citep{Engler2022}. 

\begin{figure}[h]
    \centering
    \includegraphics[width=9cm]{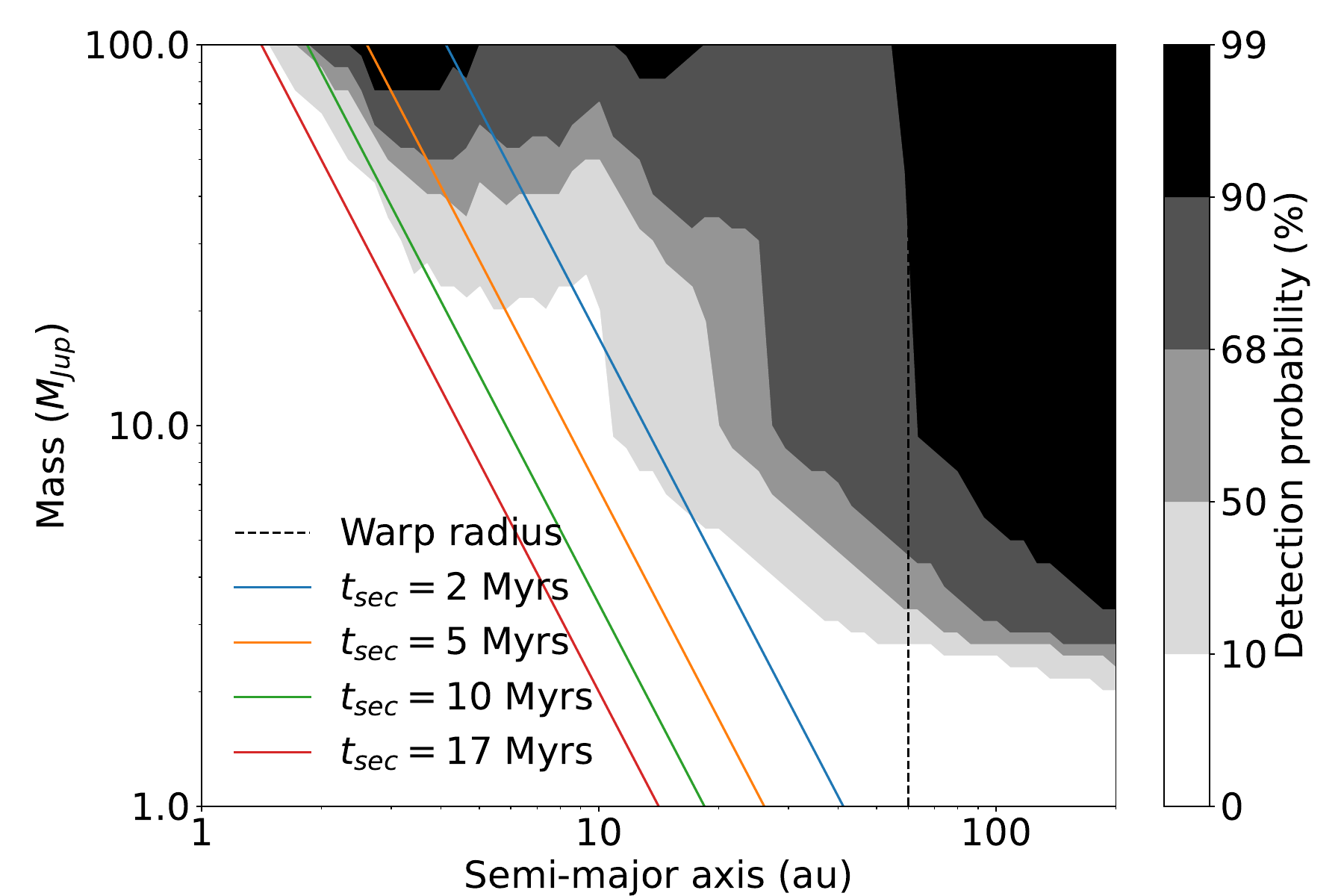}
    \caption{Sensitivity limits on bound companions, obtained by combining the constraints from IRDIS/IFS direct imaging and Gaia astrometry (proper motion anomaly and excess noise). The probability to detect a planet of a given mass and semi-major axis is colour-coded with the shade of the map; the darkest area indicates a probability greater than 90\%. The mass vs semi-major axis given in Eq.~\ref{eq:tsec} applied for a small body located at $a = 60$\,au (the warp radius) and for various timescales $t_\text{sec}$ is shown with coloured lines.}
    \label{fig:mass_sensitivity}
\end{figure}

The MESS tool generates a population of planets with orbital parameters and masses drawn from prior distributions, and tests their detectability. For the generation of orbits of planetary companions, we assumed inclinations distributed between $70^\circ$ and $90^\circ$ and eccentricities between 0 and 0.9. The result is marginalised against the planet mass and semi-major axis, and shown as a probability map in Fig.~\ref{fig:mass_sensitivity}.

We can rule out with 90\% confidence the presence of giants planets more massive than 10\,M$_\text{Jup}$ beyond 60\,au. Our constraints indicate that we are more than 50\% likely to detect any planet on an orbit beyond 20\,au and with a mass larger than 10\,M$_\text{Jup}$, for example, a \bpic{}\,b analogue orbiting at twice the separation. Below 20\,au, we are mostly blind to any planetary-mass companion or brown dwarf below 40\,$\mbox{M}_\text{Jup}$.

In Fig.~\ref{fig:mass_sensitivity} we overplot the mass of a companion that could create a warped inner disk at 60\,au after various timescales from 2\,Myr to the age of the system, 17\,Myr, using Eq.~\ref{eq:tsec}. These lines are mostly located in regions where GAIA and SPHERE are not sensitive, which calls for further investigation of this system with techniques more sensitive to giant companions with orbits in the range of $2\mbox{--}20$\,au.

%
%-------------------------------------------------------------------
\section{Conclusions}
\label{sec:concl}

%recap of main results
Using data from VLT/SPHERE and HST/STIS, we performed an in-depth analysis of the disk around \HD{}, exploring the disk morphology, vertical structure, and spectral characteristics. Our findings are as follows:
\begin{enumerate}
    \item{We traced the disk in scattered light between 20\,au and 150\,au, measuring a PA of $159.6^\circ \pm 0.6^\circ$. The surface brightness of the disk is compatible with the scenario of a wide radial belt extending up to $\sim$\,70\,au, with a peak at $35\mbox{--}40$\,au.}
    \item{The system is seen close to edge-on in the SPHERE observations; however, with STIS we see an outer wing tilt, characteristic of disks with forward scattering seen slightly inclined from edge-on. The tilt angle of $5.4^\circ \pm 0.6^\circ$ suggests the disk likely has an inclination of $80^\circ\mbox{--}84^\circ$, which is compatible with ALMA observations.}
    \item{The inner region of the disk is misaligned with the outer parts up to a separation of 60\,au, with a PA offset of $3.4^\circ \pm 0.9^\circ$ with respect to the outer disk. The perturbed inner morphology suggests the warp has not yet propagated to the outer parts of the disk, favouring the explanation of an inner perturber as the origin of the warp, as is the case for \bpic{}.}
    \item{The reflectance spectrum of the disk appears featureless, and suggests a neutral to red dust colour.}
    \item{In scattered light, the disk appears vertically thin, which is contrary to the relatively large vertical aspect ratio found in thermal emission with ALMA. This could be explained by the presence of gas in the disk, which causes small dust grains to settle to the mid-plane, while leaving larger particles unaffected. An additional explanation could be the effect of the warped inner disk, which would likely be at a different inclination than the unwarped component, and thus increasing the projected scale height in the ALMA observations where the different components are not resolved.}
    \item{Dynamical modelling supports the hypothesis of a massive perturber dynamically shaping the disk, with the orbit of the perturber being significantly inclined from, and inside, the disk.}
    \item{Using MESS3, we are able to rule out the presence of a massive companion above 10\,M$_\text{Jup}$ beyond 60\,au, and have over a 50\% likelihood of detecting a companion with a mass larger than 10\,M$_\text{Jup}$ beyond 20\,au. Due to the distance of \HD{} and the limitations of SPHERE, we cannot probe the region inside 20\,au to look for companions.}  
    
\end{enumerate}

Higher-resolution observations with ALMA could help constrain the mutual inclination between the disk and perturber, and more extensive dynamical modelling could further constrain the system. Additionally, observations using JWST, with its relatively small inner working angle and high sensitivity at large separations, could provide a view of both the inner and outer parts of the disk.

%
%-------------------------------------------------------------------
\begin{acknowledgements}

This work is based on observations made with ESO Telescopes at the Paranal Observatory under programmes 095.C-0389(A) and 095.C-0607(A).
This work has made use of the SPHERE Data Centre, jointly operated by OSUG/IPAG (Grenoble), PYTHEAS/LAM/CESAM (Marseille), OCA/Lagrange (Nice), Observatoire de Paris/LESIA (Paris), and Observatoire de Lyon.

This research has made use of the NASA Exoplanet Archive, which is operated by the California Institute of Technology, under contract with the National Aeronautics and Space Administration under the Exoplanet Exploration Program.

This project received funding from the European Research Council (ERC) under the European Union's Horizon 2020 research and innovation programme (COBREX; grant agreement n° 885593). 
This work has been supported by a grant from Labex OSUG (Investissements d’avenir – ANR10 LABX56). 
SS and JMi thank Nicolas Cuello for valuable discussion on the interpretation of the disk warp, and Sebastien Marino and Antonio Hales for sharing the radiative transfer disk model best fitting the ALMA observations of the system.

%[EC]:
This work is based on observations made with the NASA/ESA Hubble Space Telescope, obtained at the Space Telescope Science Institute, which is operated by the Association of Universities for Research in Astronomy, Inc., under NASA contract NAS5-26555. These observations are associated with program GO-15218. Support for program GO-15218 was provided by NASA through a grant from the Space Telescope Science Institute.

VF acknowledges funding from the National Aeronautics and Space Administration through the Exoplanet Research Program under Grant No. 80NSSC21K0394 (PI: S.~Ertel).
FM acknowledges funding from the European Research Council (ERC) under the European Union's Horizon 2020 research and innovation program (grant agreement No. 101053020, project Dust2Planets).

C.D. acknowledges support from the European Research Council under the European Union’s Horizon 2020 research and innovation program under grant agreement No. 832428-Origins.

In addition to the software already cited in the paper, we acknowledge the use of the following python packages: Numpy \citep{Harris2020_numpy}, Scipy \citep{Virtanen2020_scipy}; Astropy \citep{Astropy2022}, Matplotlib \citep{Hunter2007_matplotlib}; OpenCV \citep{Bradski2008_opencv}, and Photutils \citep{Bradley2022_photutils}.

\end{acknowledgements}

%-------------------------------------------------------------------
\bibliographystyle{aa}
\bibliography{hd110058_bibliography}
\nocite{*}

%-------------------------------------------------------------------
%\newpage
%\appendixpage
\begin{appendix}
\section{Gaussian fitting of disk profiles}
\label{appendix:Gauss_fit}

For the process of measuring the spine, the reduced image is rotated by an initial PA estimate so that the disk lies along the horizontal axis and divided into vertical slices, which are median combined in the horizontal direction. At larger separations, where the disk is fainter, the width of the vertical bins is increased to increase the S/N of the disk. Parameter values are provided in Table~\ref{tab:gauss_fit_params}. A Gaussian is fit to each vertical profile, with the peak position providing the central position of the spine. The standard deviation and amplitude of the Gaussian profile are used in deriving the vertical scale height and surface brightness, respectively, of the disk. The data were weighted before the fit by the standard deviation of, for SPHERE, the 'diskless' noise image in the same vertical profile, and for STIS, the upper and lower areas of the vertical profile which are free from disk signal, and the error in the fit computed from its covariance matrix.

Once the spine position was obtained, the PA was measured using the process described in \cite{Lagrange2012}. A least squares straight line fit was performed on the error weighted spine position and the slope measured. The fitting process was repeated iteratively using the rotational offset of the previous straight line fit to adjust the initial PA rotation prior to the Gaussian fit, until a null slope was obtained to give a final PA measurement and spine fit.

\begin{table}[ht]
\caption{Summery of spine fitting parameters.}
\label{tab:gauss_fit_params}
\begin{center}
\begin{tabular}{llllll}
\hline\hline
 & Inner & Outer & Bright & Bin & Init. rot.\\
 & radius & radius & radius$^{(a)}$ & size$^{(b)}$ & angle$^{(c)}$ \\
 & (px) & (px) & (px) & (px) & ($^\circ$) \\
\hline
STIS & 8 & 23 & 11 & 1 (3) & 70.0 \\
IFS & 28 & 67 & 53 & 5 (7) & 65.0 (70.4) \\
IRDI & 17 & 41 & $/$ & 3 & 65.0 (70.4) \\
\hline
\end{tabular}
\end{center}
\tablefoot{Values used to determine the vertical profiles that are median combined in the horizontal direction for bin sizes $>1$ pixel. $^{(a)}$Separation after which the bin value is increased by 2 pixels, for example, STIS bins are located at separations of [8,9,10,11,14,17,...] pixels from the image centre. $^{(b)}$Width of the vertical slices; value in the brackets are the bin size after the bright radius. $^{(c)}$Initial angle used to rotate the image clockwise so the disk lies horizontal; value in brackets are the fixed angle used for the final SPHERE spine fit derived from the STIS PA.}
\end{table}
\subsection{STIS data}
\label{appendix:STIS_fit}

The STIS image was rotated by 70.0$^\circ$ clockwise, corresponding to an initial PA estimate of 160.0$^\circ$. The vertical profile width was set at 1 pixel for the brightest parts of the disk, and 3 pixels at separations larger than 73\,au where the disk appears fainter. The fit error was combined in quadrature with an additional 0.15 pixel error, derived from spine measurements using vertical profiles with widths between 1 and 3 pixels, in addition to shifting multiple pixel bins by a single pixel.

Tests were performed on highly inclined, forward scattering synthetic disks, generated using the \texttt{scattered\_light\_disk} and \texttt{fakedisk} VIP modules, which showed straight lines fit to the outer arms of the disk would pass through the centre of the star for a symmetric disk. The straight line fits used to determine the disk PA were therefore forced through zero when measuring each side of the disk separately. 

\subsection{SPHERE data}

For analysis of the SPHERE data, we rotated the image so that the horizontal axis was aligned with the PA measured from the STIS data in Sect.~\ref{subsec:STIS_spine} of 159.6$^\circ$. The width of the vertical profiles used for the Gaussian fit was 3 and 5 pixels for the IRDIS and IFS images respectively, which was increased by 2 pixels at separations above $\sim$\,50\,au for ADI reductions, where the S/N of the disk is lower. For the RDI reduced IRDIS data, this bin size increase was omitted. 

The data were weighted by the standard deviation of the diskless noise image in the same vertical profile, and the error of the fit computed from its covariance matrix. The fit error was combined in quadrature with an additional 0.5 pixel error derived from injected fake disk fits to account for the error induced by the ADI-PCA reduction, giving the final error of the spine position measurements.

\section{Flyby scenarios}
\label{appendix:flyby}

We used the Gaia Data Release 3 (DR3) to search for possible flybys of nearby stars. We used the analytical flyby framework developed in the tool \emph{afm-spirals}\footnote{\href{https://github.com/slinling/afm-spirals}{https://github.com/slinling/afm-spirals}} detailed in \citet{Shuai2022}. 
We restricted our search to stellar neighbours currently located within 30\,pc of \HD{}, without any additional constraint on the flyby time. Among a total of 27\,367 neighbours that are identified in Gaia DR3, there are seven with a closest approach smaller than 5\,000\,au. The closest approach of the two nearest encounters, Gaia\,DR3\,6079433474860367232 and Gaia\,DR3\,6079706737855280640, is $856 \pm 1525$\,au $1691 \pm 1357$\,au respectively. It occurred 89 and 45 thousand years ago respectively. While the Gaia DR3 RUWE is 3.40 for the former star, indicating that the source is either non-single or problematic for the GAIA astrometric solution \citep{Fabricius2021}, possibly inducing biased flyby calculations in our framework, the latter star has a RUWE of 1.02, indicating that a single-star model provides a good fit to the astrometric observations.
Although those two stars are the most likely candidates to have possibly interacted with \HD{} at the time of their closest approach, the timescales at stake to impact the disk gravitationally is far beyond the age of the system (see details in Sect.~\ref{subsec:dynamical_model}). We can can therefore confidently rule out this scenario as the origin of the warp.

\clearpage

\section{Detected point sources}
\label{appendix:astrometry}

\begin{table}[h]
\caption{Astrometry of point sources detected in cADI reductions of the IRDIS observations.}
\label{tab:astrometry}
\begin{center}
\begin{tabular}{llllll}
\hline\hline
$\#$ & E.$^{(a)}$ & Ch. & Contrast & PA & Separation \\
& & & $\times10^{-6}$ & ($^\circ$) & (mas) \\
\hline
1 & 2 & H2 & 2.39\,$\pm$\,0.35 & 253.20\,$\pm$\,0.10 & 2802\,$\pm$\,7 \\
& & H3 & 3.41\,$\pm$\,0.44 & 253.22\,$\pm$\,0.09 & 2804\,$\pm$\,7 \\
2 & 2 & H2 & 2.20\,$\pm$\,0.31 & 92.19\,$\pm$\,0.09 & 3429\,$\pm$\,8 \\
& & H3 & 2.35\,$\pm$\,0.31 & 92.34\,$\pm$\,0.08 & 3430\,$\pm$\,8 \\
3 & 1 & K1 & 91.45\,$\pm$\,4.72 & 158.12\,$\pm$\,0.08 & 6196\,$\pm$\,13 \\
& & K2 & 148.60\,$\pm$\,6.91 & 158.12\,$\pm$\,0.09 & 6204\,$\pm$\,14 \\
& 2 & H2 & 80.33\,$\pm$\,4.14 & 158.18\,$\pm$\,0.08 & 6182\,$\pm$\,14 \\
& & H3 & 93.99\,$\pm$\,5.25 & 158.18\,$\pm$\,0.08 & 6186\,$\pm$\,14 \\
\hline

\end{tabular}
\end{center}
\tablefoot{Only point-source 3 was detected in both epochs. $^{(a)}$Epochs 1 and 2 refer to 2015 April 3 and 2015 April 12, respectively.}
\end{table}

%\clearpage %if further appendices, put command before new section

\end{appendix}
\end{document}